\documentclass[fleqn,10pt]{wlscirep}
\usepackage[utf8]{inputenc}
\usepackage[T1]{fontenc}
\usepackage{bm}
\usepackage{amsmath}
\usepackage{multirow}
\usepackage{caption}
\usepackage{subcaption}
\usepackage{lineno}

\usepackage{algorithm,algpseudocode}
\usepackage{algpseudocodex}
\newcounter{algsubstate}

\usepackage{cleveref}
\usepackage{mathtools}

\title{Faster ISNet for Background Bias Mitigation on Deep Neural Networks}

\author[1,2,*]{Pedro R. A. S. Bassi}
\author[3]{Sergio Decherchi}
\author[1,3,4,*]{Andrea Cavalli}
\affil[1]{Alma Mater Studiorum - University of Bologna, Bologna, Italy}
\affil[2]{Center for Biomolecular Nanotechnologies, Istituto Italiano di Tecnologia, 73010, Arnesano (LE), Italy}
\affil[3]{Istituto Italiano di Tecnologia, Genova (GE), Italy}
\affil[4]{Centre Européen de Calcul Atomique et Moléculaire (CECAM), Ecole Polytechnique Fédérale de Lausanne (EPFL), 1015, Lausanne (VD), Switzerland}
\affil[*]{e-mail: pedro.salvadorbassi2@unibo.it, andrea.cavalli@unibo.it}

\begin{abstract}
Bias or spurious correlations in image backgrounds can impact neural networks, causing shortcut learning (Clever Hans Effect) and hampering generalization to real-world data. ISNet, a recently introduced architecture, proposed the optimization of Layer-Wise Relevance Propagation (LRP, an explanation technique) heatmaps, to mitigate the influence of backgrounds on deep classifiers. However, ISNet's training time scales linearly with the number of classes in an application. Here, we propose reformulated architectures whose training time becomes independent from this number. Additionally, we introduce a concise and model-agnostic LRP implementation. We challenge the proposed architectures using synthetic background bias, and COVID-19 detection in chest X-rays, an application that commonly presents background bias. The networks hindered background attention and shortcut learning, surpassing multiple state-of-the-art models on out-of-distribution test datasets. Representing a potentially massive training speed improvement over ISNet, the proposed architectures introduce LRP optimization into a gamut of applications that the original model cannot feasibly handle.

\end{abstract}

\begin{document}

\flushbottom
\maketitle

\thispagestyle{empty}


\section{Introduction}

Deep neural networks (DNNs) can achieve high or even super-human test accuracy in some computer vision tasks. However, sometimes such performances significantly drop when models are deployed in the real-world. Shortcut learning, or Clever Hans effect\cite{ShortcutLearning}, is a possible cause for this generalization gap\cite{ShortcutLearning}. Shortcuts, or spurious correlations, are image features that correlate with the classification labels in a training dataset, but these features are not reliably present in images drawn from data distributions other than the one that originated the training data. Shortcut learning characterizes neural networks learning decision rules that erroneously take advantage of spurious correlations\cite{ShortcutLearning}. As a result, these models perform well on standard test datasets, which are independent and identically distributed (i.i.d.) with respect to the training data. However, they have impaired generalization skills in out-of-distribution (o.o.d.) data, which commonly characterizes real-world applications. For this reason, shortcut learning is an obstacle for the wide and reliable utilization of DNNs in critical scenarios, such as medical or security-related tasks.

Background bias are shortcuts in images' backgrounds. COVID-19 detection in chest X-rays is a recent example of a critical application where background bias, and consequent shortcut learning, is common\cite{NatureCovidBias}. For a long time, most large open access COVID-19 X-ray datasets contained no (or few) images displaying other diseases or healthy subjects, considering the same sources as the COVID-19 X-rays\cite{ISNet,BrixiaSet,GitCovidSet}. Therefore, many studies employed mixed datasets to train DNNs to classify COVID-19, healthy and other diseases\cite{reviewCovid}. Here, the term mixed datasets designates databases where images representing diverse classes come from distinct sources (e.g., different hospitals and cities). These sources may introduce different background characteristics in the X-rays images. As mixed datasets associate distinct sources to diverse classes, these characteristics become background bias and prompt shortcut learning. Some DNNs classifying COVID-19 performed exceedingly well on standard i.i.d. test datasets (e.g., with accuracies close to 100\%), but latter studies demonstrated that such results were boosted and affected by shortcut learning. Accordingly, the networks performance dramatically dropped when these models were evaluated with X-rays collected from hospitals that did not contribute to their training datasets (o.o.d. testing)\cite{NatureCovidBias,ShortcutCovid,critic,ISNet}.

Layer-Wise Relevance Propagation (LRP)\cite{LRP} is a technique designed to explain DNN per-sample decisions. LRP, for one sample, whose label is predicted, creates a heatmap, namely a figure composed of a back-propagated quantity called relevance. Its magnitude indicates how much each part of the input image influenced the DNN's outputs. A DNN architecture named ISNet\cite{ISNet} has recently introduced a new use for LRP, by directly optimizing LRP heatmaps to improve a deep classifier's behavior. The model produces differentiable heatmaps during training, and feeds them to a heatmap loss\cite{ISNet}. The loss function uses segmentation masks to identify and penalize background attention in the maps. By minimizing a linear combination of the heatmap loss and a classification loss, the ISNet learns decision rules that ignore background bias. This optimization procedure is called background relevance minimization, and it can be regarded as an explanation-based spatial attention mechanism. Notably, the ISNet does not need LRP heatmaps nor segmentation masks for inference, ensuring that the run-time model has no extra computation cost in relation to a standard classifier. Thus, it can be efficiently deployed in portable or embedded devices.

The ISNet was tested on synthetic background bias (inserted into natural images), and on standard mixed tuberculosis and COVID-19 X-ray datasets. By disregarding the background in its decisions, the ISNet hindered shortcut learning and improved generalization. Thus, performance on o.o.d. test datasets surpassed the multiple state-of-the-art DNNs it was compared to. Besides empirically comparing the ISNet to several alternative methodologies, the study presenting the ISNet explained in detail the algorithmic reasons for its superior performance\cite{ISNet}. Besides the ISNet, the usual segmentation-classification pipeline was the only other network whose decisions were consistently not influenced by background bias. However, this model was less accurate, heavier and much slower in inference (needing to first run a deep segmenter, then classify the image with erased background). The ISNet's main drawback is that its training time linearly increases with the number of categories in the classification task\cite{ISNet}. Accordingly, ISNet's training computational cost becomes unfeasible for problems with many classes.


Our main contributions are: \textbf{(1)} to propose 3 deep classifier architectures, collectively named Faster ISNet. By reformulating the ISNet learning procedure, training time becomes independent of the number of classes. Hence, the models are drastically faster than the original ISNet when the number of classes is significant. Accordingly, this study feasibly introduces LRP optimization into a new gamut of applications. The inference phase is not affected by our reformulations. \textbf{(2)} To improve the ISNet loss (\cref{heatmapLoss}). \textbf{(3)} To suggest a new heuristic to accelerate ISNet's hyper-parameter search (\cref{tuneCut}). \textbf{(4)} To introduce LRP Deep Supervision, a technique to improve the ISNet convergence and background bias robustness. \textbf{(5)} To create LRP-Flex, a LRP implementation based on a reformulation of LRP's rules. Its main advantages over most LRP implementations (including ISNet's\cite{ISNet}) is conciseness and model agnosticism, not requiring architecture-specific code. Thus, LRP-Flex is a practical tool to explain arbitrary DNNs. The original ISNet was only implemented for three backbone architectures, but LRP-Flex can promptly convert any ReLU-based DNN into an ISNet or Faster ISNet. 

Here, we experiment with DenseNet\cite{DenseNet} and the popular ResNet\cite{ResNet} backbones. To quantitatively measure the Faster ISNet's background bias robustness, we challenge it with synthetic background bias, which we inserted in the MNIST\cite{mnist} and Stanford Dogs\cite{StanfordDogs} datasets. Additionally, we train the DNNs for COVID-19 detection on chest X-rays, using the mixed dataset presented in the original ISNet study\cite{ISNet}. We evaluate the DNNs with X-rays from hospitals and cities that did not contribute to the training data, assessing whether the Faster ISNet hinders background bias attention and the consequent shortcut learning, thus improving o.o.d. generalization. We compare our model to the original ISNet and the multiple state-of-the-art classifier architectures used as benchmark in the original ISNet study\cite{ISNet}. Code for the Faster ISNet and LRP-Flex is available at \url{https://github.com/PedroRASB/FasterISNet}.

\section{Methods}
\subsection{Layer-Wise Relevance Propagation}
\label{LRP}

DNNs are complex non-linear models, with possibly millions of parameters. Therefore, understanding the reasons for a DNN's decision is challenging. However, this knowledge is fundamental for critical and security related applications (such as AI-assisted diagnosis), where trustworthiness is key. Layer-Wise Relevance Propagation\cite{LRP} (LRP) is an explanation technique, i.e., a method to elucidate the reasons for a DNN output. LRP creates heatmaps, figures that indicate how each element (e.g., pixel) in the DNN input influenced the model's output. LRP backpropagates a quantity called relevance, from a chosen DNN output (logit), up to the network's input. The final relevance values constitute the heatmap, which bears the same shape as the input image. Positive relevance shows input regions that were responsible for increasing the chosen logit, i.e. they represent positive evidence for the class associated with the logit. Conversely, negative relevance indicates input areas that contributed to reducing the logit value, constituting negative evidence. Moreover, the relevance's absolute value informs how important the input region was for the DNN decision. The LRP propagation can start from any DNN logit, and the resulting heatmap will reveal positive and negative evidence for the class associated with such logit.

LRP uses semi-conservative rules to propagate the relevance layer-by-layer. In other words, the rules are designed to produce minimal destruction or creation of relevance, mostly redistributing it\cite{LRPBook}. This property ensures a strong relationship between the heatmap elements and the DNN output\cite{LRPBook}. Furthermore, the propagation procedure considers all the DNN parameters and layers' activations, embedding in a single heatmap both the late layers' high-level semantics and context information, and the early layers' high-definition spatial information\cite{ISNet}. Finally, LRP is a principled technique: the rules mentioned in this study (LRP-0, LRP-$\varepsilon$, LRP-z$^{B}$ and LRP-z$^{+}$) are justified by the Deep Taylor Decomposition (DTD) framework. Briefly, these rules propagate relevance according to a series of approximate Taylor expansions performed at the DNN neurons\cite{LRPBook}, which capture the influence that a neuron exerts on its peers. Moreover, the LRP rules are differentiable, and the technique is computationally efficient. The time required to create a heatmap is similar to the time needed for a standard gradient backward pass\cite{ISNet}.

For clarity, we explain how LRP-$\varepsilon$ and LRP-z\textsuperscript{+} propagate relevance through a fully-connected layer L with ReLU (Rectified Linear Unit) activation. However, the same equations are valid for convolutional layers followed by ReLU, as their pre-activation outputs are also linear combinations of their inputs. Moreover, the equations can also apply to batch normalization, dropout, pooling, and other common layers, by expressing them as equivalent fully-connected or convolutions layers, or fusing them with adjacent dense layers or convolutions\cite{LRPBook,ISNet}.

\Cref{dense_equation} shows a fully-connected layer's outputs, $z_{k}^{L}$ ($L$ is a layer identification superscript), before the ReLU function. $w_{jk}^{L}$ represents the weight connecting the layer's input j to its output k, $w_{0k}^{L}$ is the output k bias parameter (consider $a_{0}^{L}=1$), and $a_{j}^{L}$ is the layer's j-th input. \Cref{LRP-e_equation} represents LRP-$\varepsilon$\cite{LRP}. It redistributes the relevance at the layer L output, $R_{k}^{L+1}$ (i.e., the relevance referent to the input of layer L+1, $a_{k}^{L+1}=\textrm{ReLU}(z_{k}^{L})$), to the layer's inputs ($a_{j}^{L}$), producing their respective relevances, $R_{j}^{L}$. Here, the sign($\cdot$) function returns 1 for positive or 0 arguments, and -1 for negative ones. The $\epsilon$ hyper-parameter is a small positive constant, used to avoid division by zero, improve numerical stability, and denoise heatmaps\cite{ISNet}. If $\epsilon=0$, \cref{LRP-e_equation} represents LRP-0 instead of LRP-$\varepsilon$. The quotient in \cref{LRP-e_equation} shows that LRP redistributes relevance according to how much each input element, $a_{j}^{L}$, contributed to each layer output, $z_{k}^{L}$.  \cref{LRP-e_begin} shows the initial relevance values when explaining the logit relative to class c ($z_{c}^{\textrm{Lmax}}$, where Lmax indicates the last DNN layer).





\begin{gather}
\label{dense_equation}
z_{k}^{L}=\sum_{j} w_{jk}^{L}a_{j}^{L}\\
\label{LRP-e_equation}
R_{j}^{L}=\sum_{k}\frac{w_{jk}^{L}a_{j}^{L}}{z_{k}^{L}+\textrm{sign}(z_{k}^{L})\epsilon}R_{k}^{L+1}\\
\label{LRP-e_begin}
R_{k}^{\textrm{Lmax}+1}=
\begin{cases}
    z_{c}^{\textrm{Lmax}} \mbox{ if } k=c\\
    0 \mbox{ otherwise}
\end{cases}
\end{gather}



The LRP-z\textsuperscript{+} rule\cite{LRPZb} (\cref{lrp-z+_equation}) is similar to LRP-$\varepsilon$, but it only propagates positive relevance. Thus, it ignores the negative evidence for a DNN decision. The superscript + indicates the maximum between a value and zero ($x^{+}=\textrm{max}(x,0)$). Originally, $\epsilon$ was not present in LRP-z\textsuperscript{+}. We included it for numerical stability during Dual ISNet training (see later). \Cref{LRP-z+_begin} shows the beginning of the relevance propagation when explaining class c logit with LRP-z\textsuperscript{+}\cite{LRPBook}. 

\begin{gather}
\label{lrp-z+_equation}
R_{j}^{L}=\sum_{k}\frac{(w_{jk}^{L}a_{j}^{L})^{+}}{\sum_{j} (w_{jk}^{L}a_{j}^{L})^{+}+\epsilon}R_{k}^{L+1} \\
\label{LRP-z+_begin}
R_{k}^{\textrm{Lmax}+1}=
\begin{cases}
    \sum_{j} (w_{jc}^{\textrm{Lmax}}a_{j}^{\textrm{Lmax}})^{+} \mbox{ if } k=c\\
    0 \mbox{ otherwise}
\end{cases}
\end{gather}

\subsection{Original ISNet}
\label{isnetOriginal}


Layer-Wise Relevance Propagation's goal is to interpret deep neural networks. The direct optimization of LRP heatmaps to improve a deep classifier's attention was recently proposed in the ISNet paper\cite{ISNet}. The architecture is defined by a training procedure named Background Relevance Minimization (BRM), which minimizes the background relevance magnitude in a classifier's LRP heatmaps. The ISNet Loss ($L_{IS}$) is a linear combination of two functions, a standard classification loss (e.g., cross-entropy) $L_{C}$ and a heatmap loss $L_{LRP}$. Globally one has $L_{IS} = (1-P) L_{C}+P L_{LRP}$, where the hyper-parameter $P \in (0,1)$ rules the trade-off between background attention rejection and regular classification loss fitting. $L_{LRP}$ (\cref{heatmapLoss}) utilizes training time known segmentation masks to identify the background in the classifier's LRP heatmaps, and it penalizes background relevance. The masks are images valued 1 over the foreground, and 0 in the background. They may be manually drawn or produced by pre-trained semantic segmenters, which can be application-specific (e.g., U-Net\cite{unet}), or general novel class segmentation DNNs (e.g., DeepMAC\cite{deepMAC}). The heatmap loss confers to the network the ability to work correctly even on o.o.d data at run-time, as the network learns to ignore the background.

During the training phase, the main difference with respect to a traditional DNN is that in order to evaluate the second loss one needs to create LRP heatmaps of the current batch. In detail one employs LRP-$\varepsilon$ throughout the DNN, except for the first layer, which utilizes LRP-z$^{\mathrm{B}}$. Due to LRP-$\varepsilon$ denoising properties, the ISNet Loss could stably and efficiently converge even for very deep backbones, unlike what was observed in Gradient*Input optimization\cite{ISNet}. At inference time, the ISNet classifier can focus exclusively on the image foreground without the help of segmentation masks nor any auxiliary semantic segmentation DNN. The consequence is that the inference ISNet's architecture and computational cost are identical to a standard classifier's. However, the network's decision rules rely only on the image's foreground features, indicating the network acquired an Implicit Segmentation skill, from which the name ISNet. The precision of the ISNet's Implicit Segmentation\cite{ISNet} is justified by the high level of abstraction and precise spatial information portrayed in LRP heatmaps. Moreover, since LRP captures how DNN neurons influence their subsequent peers (using approximate Taylor expansions), the minimization of the LRP relevance flow to the images' backgrounds constricts the corresponding influence flow from background bias to the DNN's output, justifying why the ISNet's decisions are robust to the effect of background bias\cite{ISNet}.


\subsection{Faster ISNet}

When one starts the LRP propagation from a given class score (logit), the resulting heatmap explains how each input element influenced that logit. Therefore, the map may not properly capture the effect of background bias on logits representing other classes. The standard LRP procedure is to explain the highest logit, which represents the winning class\cite{LRPBook}. In this case, the resulting heatmap may not show attention to a background bias that reduced the losing classes' logits. However, this background bias can alter the classifier's decision. Accordingly, minimizing background relevance only in the winning classes heatmaps could not hinder background bias attention\cite{ISNet}.

To account for the background influence over all class scores, the original ISNet generates and optimizes, for each training image, one heatmap explaining each possible logit. The multiple heatmaps can be processed in parallel (like batch samples). However, considering a classification task with $C$ categories, ISNet's training time increases approximately linearly with $C$. Memory consumption also increases, and, when it reaches the available limit, the heatmaps need to be produced serially, increasing training time linearly with $C$\cite{ISNet}.

Here, we present three ISNet reformulations, which use alternative LRP procedures to enable the creation of only one, or two, heatmaps per training image. Therefore, the here proposed architectures remove the dependency of the training time (or memory consumption) on $C$. Accordingly, we collectively call them "Faster ISNet". We introduce three new variants: Dual ISNet, Selective ISNet and Stochastic ISNet.

\subsubsection{Dual ISNet.}
\label{DualISNet}
To avoid creating and penalizing multiple heatmaps, we should ideally produce and optimize a single LRP map that reveals how input features influence all the classifier's class scores (logits) jointly. We can propagate relevance from all logits simultaneously instead of propagating it from a single logit $z_{c}^{\textrm{Lmax}}$. Accordingly, instead of using \cref{LRP-e_begin} to define the logit's relevances, we set each logit relevance as the logit value ($R_{k}^{\textrm{Lmax+1}}=z_{k}^{\textrm{Lmax}} \forall k\in[1,2,...,C]$). Then, we propagate these relevances with the standard LRP-$\varepsilon$ rule (and LRP-z$^{B}$ in the first DNN layer, optionally, \cref{TrainingProcedure}). We dubbed the resulting explanation the joint LRP-$\varepsilon$ heatmap. Creating a joint heatmap takes the same time as creating a standard map. Unfortunately, during the relevance propagation, positive background relevance originating from a logit may encounter negative background relevance that originated from another logit, causing destructive interference in the joint LRP-$\varepsilon$ heatmap. This phenomenon can deceivingly reduce the amount of background relevance in the joint heatmap. Hence, it is not reliable to solely base the ISNet Loss on joint LRP-$\varepsilon$ heatmaps.

As the LRP-z$^{+}$ rule only propagates positive relevance, the joint LRP-z$^{+}$ map is not subject to destructive interference. To produce this heatmap, we set the logits' relevances as $R_{k}^{\textrm{Lmax}+1}=\sum_{j} (w_{jk}^{\textrm{Lmax}}a_{j}^{\textrm{Lmax}})^{+} \forall k\in[1,2,...,C]$, instead of using \cref{LRP-z+_begin}. Then, we further propagate the relevance with the standard LRP-z$^{+}$ rule (\cref{lrp-z+_equation}), except for the first DNN layer, where may use LRP-z$^{B}$. Optimizing solely the joint LRP-z$^{+}$ map can also be insufficient to minimize the influence of background bias on the classifier. By not propagating negative relevance, the LRP-z$^{+}$ explanation ignores negative evidence (\cref{LRP}), which may also cause shortcut learning\cite{ISNet}. 

To ensure background bias resistance, the Dual ISNet creates two LRP heatmaps per training image: the LRP-$\varepsilon$ joint map, and the LRP-z$^{+}$ joint map. Both are individually penalized by the heatmap loss. As LRP-z$^{+}$ is immune to destructive interference, the background relevance minimization on the LRP-z$^{+}$ joint map will not allow the image background to positively contribute to the DNN logits. Consequently, LRP-z$^{+}$ optimization also minimizes positive background relevance in the joint LRP-$\varepsilon$ heatmap, avoiding destructive interference in the map's background. Accordingly, background relevance minimization in the LRP-$\varepsilon$ map becomes able to hinder negative evidence in the image's background. Therefore, the Dual ISNet minimizes the influence of the background on the classifier decisions. Moreover, it introduces a potentially massive training speed improvement, by replacing the C (number of classes) LRP-$\varepsilon$ heatmaps in the Original ISNet by only two joint heatmaps. \Cref{lrpBlock} explains a fast procedure to compute LRP-z$^{+}$ and LRP-$\varepsilon$ joint heatmaps, which we included in the LRP Block (the ISNet LRP implementation).

\subsubsection{Selective ISNet.}
\label{SelectiveISNet}
The Selective ISNet produces and optimizes one LRP heatmap per training sample. To ensure background bias resistance, the map must capture the influence of the image background over all logits. Thus, the Selective ISNet is defined by the following reformulation of the ISNet LRP procedure: instead of creating C LRP heatmaps to explain the C DNN logits ($z_{c}^{\textrm{Lmax}}$, where $c\in[1,C]$), create a single LRP heatmap that explains a Softmax-based quantity $\eta_{c}$ (\cref{nFrac}), where c is the class associated to the lowest DNN logit. 

\begin{gather}
\label{nFrac}
\eta_{c}=\ln(\frac{P_{c}}{1-P_{c}})\\
\label{softmax}
\mbox{where: } P_{c}=\frac{e^{z_{c}^{\textrm{Lmax}}}}{\sum_{c'=1}^{C}e^{z_{c'}^{\textrm{Lmax}}}}
\end{gather}

$\eta_{c}$ is a monotonically increasing function of $P_{c}$, the Softmax predicted probability of class c (\cref{softmax}). $P_{c}$ depends on all DNN's logits. Thus, the heatmap explaining $\eta_{c}$ depends on how the input features affect all class scores; e.g., a background bias that only reduces the losing classes' logits will consequently increase the classifier confidence for the winning class ($P_{h}$). Thus, the bias will increase $\eta_{h}$. Accordingly, it produce positive background relevance in the LRP heatmap explaining $\eta_{h}$, even though the bias does not appear in the heatmap explaining the logit $z_{h}^{\textrm{Lmax}}$. The explanation of $\eta_{c}$ instead of logits was originally proposed to make LRP heatmaps more class-selective\cite{LRPBook}. Indeed, $\eta_{c}$ heatmaps better reveal how input features affected the classifier confidence for class c ($P_{c}$). For example, in a heatmap explaining a logit $z_{c}^{\textrm{Lmax}}$, an input feature can be represented as positive relevance for having increased $z_{c}^{\textrm{Lmax}}$. However, the feature may have actually reduced $P_{c}$, by strongly incrementing the other logits. In this case, the feature will reduce $\eta_{c}$, and be represented as negative relevance in the $\eta_{c}$ heatmap. 

To explain $\eta_{c}$ instead of the logit $z_{c}^{\textrm{Lmax}}$, we only change the LRP propagation rule for the last DNN layer. Instead of using the procedure in \cref{LRP}, we use \cref{selectiveEquation1} to obtain the LRP relevance at the input of the last layer ($R_{j}$)\cite{LRPBook}. Then, the relevance is further propagated with LRP-$\varepsilon$ (and LRP-z$^{B}$ in the input layer, optionally, \cref{TrainingProcedure}). In the equation, $a_{j}^{\textrm{Lmax}}$ represents the DNN last layer inputs, $w_{jc}^{\textrm{Lmax}}$ is the weight connecting input $a_{j}^{\textrm{Lmax}}$ to logit $z_{c}^{\textrm{Lmax}}$ ($w_{0c}^{\textrm{Lmax}}$ is a bias parameter, and $a_{0c}^{\textrm{Lmax}}=1$), $\epsilon$ is the LRP-$\epsilon$ stabilizer hyper-parameter (set to $10^{-2}$), and $\mu$ is a small positive constant ($10^{-5}$), which we add to improve numerical stability. The equation is defined for a fully-connected last layer, but it is also applicable to other layers that can be expressed as a fully-connected layer (e.g., convolution). \Cref{lrpBlock} shows how we altered the ISNet LRP implementation (the LRP Block) to apply \cref{selectiveEquation1} in an computationally convenient manner.

\begin{gather}
\label{selectiveEquation1}
R_{j}^{\textrm{Lmax}}=\sum_{c'=1}^{C}\frac{(w_{jc}^{\textrm{Lmax}}-w_{jc'}^{\textrm{Lmax}})a_{j}^{\textrm{Lmax}}}{z_{c,c'}+\textrm{sign}(z_{c,c'})\epsilon}R_{c,c'} \\
\label{selectiveEquation3}
 \mbox{where: } R_{c,c'}=\frac{z_{c,c'} e^{-z_{c,c'}}}{\sum_{c''\neq c}e^{-z_{c,c''}}+\mu}\\
\label{selectiveEquation2}
z_{c,c'}=z_{c}^{\textrm{Lmax}}-z_{c'}^{\textrm{Lmax}}=\sum_{j}a_{j}^{\textrm{Lmax}}(w_{jc}^{\textrm{Lmax}}-w_{jc'}^{\textrm{Lmax}})
\end{gather}

$\eta_{c}$ is an unbounded quantity, whose magnitude increases with the difference between $P_{c}$ and 0.5, where $\eta_{c}=0$. Thus, when we start the LRP procedure from $\eta_{c}$, small and high $P_{c}$ normally produce heatmaps containing higher relevance magnitude. Indeed, $P_{c}=0.5$ represents a state of maximal classifier uncertainty for class c (50\% probability), justifying why LRP relevances should be closer to zero. During the beginning of the training procedure, it is common for the highest class probability ($P_{c}$) to be near 50\%. However, it is rarer for the lowest $P_{c}$ to be near 0.5, especially with multiple possible classes. Therefore, to create more expressive heatmaps, we chose to explain $\eta_{c}$ for the class c corresponding to the lowest DNN logit (for each training image).

\subsubsection{Stochastic ISNet.}
\label{StochasticISNet}
As previously discussed, the heatmap explaining a single class logit cannot properly capture how bias affects all class scores. Thus, the standard practice of producing LRP heatmaps for the winning class or the label class\cite{LRPBook} is not adequate for background relevance minimization. However, we hypothesize that the minimization of background relevance in random class (c) heatmaps may avoid background bias attention. 

A trivial strategy would be choosing c from an uniform probability distribution for each training image, then creating and optimizing the LRP-$\varepsilon$ map that explains logit c. However, assigning the same probabilities for all classes could be problematic. Only the map explaining the highest logit can properly show positive correlations between the winning logit and the presence of background bias. Such correlations can strongly affect the classifier decisions and must be effectively penalized. Let us consider an unbalanced classification dataset with C categories, where class A has a small number of samples, N. Moreover, assume that the classifier being trained already has high accuracy. In this case, following the random uniform selection of logits, only about N/C winning logit heatmaps (explaining class A) will be created for the class A images in one epoch. If N is small and C is large, a positive correlation between background bias and the class A logit will have a small effect in the average heatmap loss. Thus, background bias attention will not be effectively hindered for class A. For this reason, we devised the following logit selection strategy: give the highest logit 50\% selection probability, while all other logits have (50/(C-1))\% probability. 

Using this strategy, the Stochastic ISNet selects one logit per training image and explains it with a LRP-$\varepsilon$ heatmap (using LRP-z$^{B}$ in the first DNN layer, optionally, \cref{TrainingProcedure}), which the heatmap loss optimizes. During training, the technique creates approximately the same number of winning and losing class heatmaps. As previously explained, the optimization of winning class heatmaps are important to hinder positive correlations between background bias and the winning logits. Meanwhile, the penalization of losing class heatmaps prevents the background bias from acting as negative evidence and influencing the classifier decision by reducing the losing class logits. The Stochastic ISNet requires no modifications to the original ISNet LRP implementation (LRP Block). 

\subsection{LRP-Flex: a Simple, Fast and Model-Agnostic Implementation of LRP}
\label{Flex}

Layer-Wise Relevance Propagation can explain virtually any DNN architecture\cite{LRPBook}. However, specific coding is normally required for each different architecture. Moreover, LRP libraries are commonly large and complex, especially when implementing LRP for multiple architectures. The original ISNet LRP Block\cite{ISNet} implemented the rules LRP-$\varepsilon$ and LRP-z$^{B}$ for the DenseNet\cite{DenseNet}, VGG\cite{VGG} and simple sequential networks (defined as a PyTorch Sequential object), using about 2000 lines of code. With our inclusion of the LRP-z$^{+}$ rule and the LRP ResNet implementation, it now has about 4000 lines.

We introduce LRP-Flex, an implementation of LRP-$\varepsilon$ for DNNs utilizing only ReLU nonlinearities in their hidden layers (the last layer can have alternative activations). Its key features are: first, it is exceedingly simple, requiring significantly less code lines than standard LRP implementations (e.g., the LRP-Flex PyTorch code is about 10 times shorter than the LRP Block); second, it is model-agnostic, being readily applicable to arbitrary DNN architectures, and not requiring the user to spend time writing architecture-specific code; third, it is fast, taking advantage of highly optimized backpropagation engines available in deep learning libraries. LRP-Flex produces differentiable heatmaps. Thus, it can be employed to easily implement the original ISNet, the Stochastic ISNet, and the Selective ISNet for any ReLU-based classifier architecture. Furthermore, it is a practical and fast technique to explain DNN's decisions with LRP. We summarize LRP-Flex workflow below. The algorithm is based on an equivalent reformulation of LRP, which is elucidated in \cref{FlexAppendix}. The workflow considers the LRP-$\varepsilon$ rule, but it can be easily expanded to use other rules in specific DNN layers (\cref{FlexAppendix}); e.g., we may use LRP-z$^{B}$ for the first DNN layer.

\begin{enumerate}
    \item Initialization: modify the gradient backpropagation procedure for all ReLU functions in the neural network (e.g., using PyTorch's backward hooks). The equation below defines the modified ReLU backpropagation rule. $\mathbf{G^{out}}$ (with elements $G^{out}_{k}$) is the quantity that was back-propagated until the output of the ReLU function ($\mathbf{a^{out}}$, with elements $a^{out}_{k}$). The equation back-propagates $\mathbf{G^{out}}$ to the input of the ReLU, producing $\mathbf{G^{in}}$ (with elements $G^{in}_{j}$). The variable $\varepsilon$ is the small positive hyper-parameter in LRP-$\varepsilon$ (e.g., 0.01). This backward pass modification has to be deactivated when not creating LRP heatmaps (e.g., for loss gradient calculation). \\
    $G_{k}^{in} = \frac{a_{k}^{out}}{a_{k}^{out}+
    \varepsilon} G_{k}^{out}$
    
    \item Forward pass: run the neural network and store the outputs of its ReLU functions (e.g., employing forward hooks in PyTorch).
    
    \item Modified backward pass: to explain the DNN logit $z_{c}^{\textrm{Lmax}}$ (or $\eta_{c}$, for the Selective ISNet), request the automatic backpropagation engine in the deep learning library to calculate the gradient of $z_{c}^{\textrm{Lmax}}$ (or $\tilde{\eta_{c}}$) with respect to the network's input ($\mathbf{X}$). However, use the modified backward procedure in all ReLU functions (Step 1). Accordingly, the resulting tensor ($\mathbf{G^{0}}$) will not match the actual input gradient, $\mathbf{G^{0}} \neq \nabla_{\mathbf{X}}z_{c}^{\textrm{Lmax}}$ (or $\mathbf{G^{0}} \neq \nabla_{\mathbf{X}}\tilde{\eta_{c}}$). $\tilde{\eta_{c}}$, defined below, is a bounded version of $\eta_{c}$ (\cref{nFrac}). $\mu$ is a small positive hyper-parameter (e.g., 0.01), which ensures numerical stability, and $P_{c}$ is the softmax-estimated class c probability (\cref{softmax}).\\
    $\tilde{\eta_{c}}=\ln(P_{c}+\mu)-\ln(1-P_{c}+\mu)$
    
    \item Element-wise multiplication: to obtain the final LRP-$\varepsilon$ heatmap ($\mathbf{R^{0}}$), element-wise multiply the back-propagated quantity ($\mathbf{G^{0}}$) and the DNN input ($\mathbf{X}$).\\
    $\mathbf{R^{0}}=\mathbf{G^{0}}\odot \mathbf{X}$
\end{enumerate}

\subsection{LRP Deep Supervision}

During relevance propagation, LRP produces intermediate heatmaps at each DNN depth. They explain how hidden layers' inputs influenced the DNN output. Since all these heatmaps portray a high level of abstraction (which LRP carries from deep layers), we are able to minimize their background relevance. Here, we introduce LRP Deep Supervision (LDS), a technique that leverages intermediate LRP heatmaps to improve the ISNet convergence. It resizes foreground segmentation masks to fit multiple intermediate heatmaps and applies the ISNet's heatmap loss to them. All losses use the same hyper-parameters, except for $C_{1}$ and $C_{2}$, which we automatically set per-layer (\cref{tuneCut}). To aggregate the multiple losses into a scalar loss, we use Global Weighted Ranked Pooling (GWRP): we arrange the losses in descending order, give them exponentially decreasing weights, and perform a weighted average\cite{GWRP}. Thus, GWRP prioritizes high losses, ensuring none of the supervised heatmaps has significant background attention. Intermediate heatmap losses are auxiliary objectives to improve the convergence of the standard (input-level) heatmap loss. Globally, LDS constrains the DNN to discard information from both the input's and the feature maps' backgrounds. Thus, it enforces direct mapping, encouraging feature maps' foregrounds to portray only relevant information, extracted solely from the DNN input's foreground. The L\textsuperscript{th} layer's LRP heatmap expresses the attention of a sub-network, defined by layer L and all subsequent layers. Locally, LDS guides multiple sub-networks to focus on the foregrounds of their input feature maps. Optimizing the attention of a smaller sub-network should be easier than optimizing the entire DNN's focus. Moreover, the improvement of a sub-network’s attention profile should help the optimization of larger sub-networks. Hence, we expect LDS to improve ISNet loss convergence in difficult optimization problems, increasing robustness and accuracy. To reduce computational cost, LDS supervises only a few DNN layers. We prioritize layers representing signal bottlenecks, i.e., layers that are not in parallel with skip connections (\cref{TrainingProcedure}).

\section{Results}

\begin{table}[!h]
\centering
\caption{Summary of the Test Results for the 3 Applications}
\label{AllResults}
\begin{tabular}{|l|l|l|l|l|l|l|l|} 
\hline
-                        & \multicolumn{3}{l|}{\begin{tabular}[c]{@{}l@{}}Synthetically\\Biased MNIST\end{tabular}}                                                                           & \multicolumn{3}{l|}{\begin{tabular}[c]{@{}l@{}}Synthetically Biased\\Stanford Dogs\end{tabular}}                                                                   & \begin{tabular}[c]{@{}l@{}}COVID-19\\Detection\end{tabular}  \\ 
\hline
Neural Network           & \begin{tabular}[c]{@{}l@{}}ACC\\i.i.d.\end{tabular} & \begin{tabular}[c]{@{}l@{}}ACC\\o.o.d.\end{tabular} & \begin{tabular}[c]{@{}l@{}}ACC\\Deceiving\end{tabular} & \begin{tabular}[c]{@{}l@{}}AUC\\i.i.d.\end{tabular} & \begin{tabular}[c]{@{}l@{}}AUC\\o.o.d.\end{tabular} & \begin{tabular}[c]{@{}l@{}}AUC\\Deceiving\end{tabular} & \begin{tabular}[c]{@{}l@{}}maF1\\o.o.d.\end{tabular}         \\ 
\hline
Standard Classifier      & 1                                                   & 0.654                                               & 0.124                                                  & 1                                                   & 0.547                                               & 0.454                                                  & 0.546$\pm$0.01                                               \\ 
\hline
Dual ISNet               & 0.979                                               & 0.979                                               & 0.979                                                  & 0.905                                               & 0.905                                               & 0.905                                                  & 0.715$\pm$0.009                                              \\ 
\hline
Selective ISNet          & 0.967                                               & 0.967                                               & 0.967                                                  & 0.799                                               & 0.798                                               & 0.799                                                  & 0.77$\pm$0.008                                               \\ 
\hline
Stochastic ISNet         & 0.983                                               & 0.983                                               & 0.983                                                  & 0.859                                               & 0.857                                               & 0.857                                                  & 0.731$\pm$0.009                                              \\ 
\hline
ISNet Softmax Grad*Input & 0.949                                               & 0.949                                               & 0.949                                                  & 0.55                                                & 0.55                                                & 0.55                                                   & 0.323$\pm$0.007                                              \\ 
\hline
Original ISNet           & 0.985                                               & 0.985                                               & 0.985                                                  & -                                                   & -                                                   & -                                                      & 0.773$\pm$0.009                                              \\ 
\hline
RRR                      & 0.97                                                & 0.967                                               & 0.967                                                  & 0.941                                               & 0.569                                               & 0.465                                                  & 0.55$\pm$0.009                                               \\ 
\hline
Reference Classifier     & -                                                   & 0.985                                               & -                                                      & -                                                   & 0.809                                               & -                                                      & -                                                            \\ 
\hline
U-Net+Classifier         & -                                                   & -                                                   & -                                                      & 0.896                                               & 0.897                                               & 0.897                                                  & 0.645$\pm$0.009                                              \\ 
\hline
Multi-task U-Net         & -                                                   & -                                                   & -                                                      & 0.972                                               & 0.824                                               & 0.631                                                  & 0.374$\pm$0.01                                               \\ 
\hline
AG-Sononet               & -                                                   & -                                                   & -                                                      & 1                                                   & 0.507                                               & 0.419                                                  & 0.356$\pm$0.008                                              \\ 
\hline
Extended GAIN            & -                                                   & -                                                   & -                                                      & 0.804                                               & 0.803                                               & 0.803                                                  & 0.466$\pm$0.009                                              \\ 
\hline
Vision Transformer       & -                                                   & -                                                   & -                                                      & 1                                                   & 0.578                                               & 0.46                                                   & 0.46$\pm$0.009                                               \\ 
\hline
Dual ISNet LDS           & -                                                   & -                                                   & -                                                      & 0.895                                               & 0.895                                               & 0.895                                                  & -                                                            \\ 
\hline
Selective ISNet LDS      & -                                                   & -                                                   & -                                                      & 0.9                                                 & 0.9                                                 & 0.9                                                    & -                                                            \\ 
\hline
Stochastic ISNet LDS     & -                                                   & -                                                   & -                                                      & 0.885                                               & 0.885                                               & 0.885                                                  & -                                                            \\
\hline
\end{tabular}
\end{table}

Datasets and their limitations are detailed in \cref{datasets}, data processing in \ref{DataProcessing}, benchmark architectures in \ref{alternative}, and training procedure in \ref{TrainingProcedure}. \Cref{AllResults} reports results on 3 applications. We begin with synthetically biased MNIST\cite{mnist}, where training images contain white background pixels whose positions correlate to the image classes (digits). In this application, DNNs use a ResNet18 backbone. We also implemented a reference classifier (ResNet18), a DNN trained without synthetic bias. In synthetic bias applications, we considered 3 versions of the test set: i.i.d., containing the synthetic bias; o.o.d., with no bias; and deceiving bias, where we changed the correlations between biases and image classes. The test accuracy reduction when synthetic bias is removed or replaced by confounding bias is a quantitative measurement of background bias influence on DNNs, quantifying shortcut learning. The standard classifier (ResNet18) reveals the extreme tendency for shortcut learning in the dataset: its accuracy drops from 100\% to 65.4\% and to 12.4\% when the bias is removed or substituted by confounding bias, respectively. RRR\cite{RRR} was fairly robust, with 0.3\% accuracy drop upon bias removal. All ISNet variants had similar results and were resistant to background bias, displaying no accuracy drop across the 3 test settings. Also, they were accurate, performing similarly to the DNN trained without bias.

\Cref{AllResults} also reports results (macro averaged one-vs-one ROC-AUC\cite{MulticlassAUC}) for the synthetically biased Stanford Dogs dataset. It presents a difficult fine-grained classification problem, with 120 dog breeds (classes), small inter-class variation, and large intra-class variation\cite{StanfordDogs}. The synthetic bias is a number representing the dog breed, added to the image's background. Here, all DNNs use a ResNet50 backbone, except for vision transformer\cite{VisionTransformer}, Multi-task U-Net\cite{ISNet} and AG-Sononet\cite{AGNet} (\cref{alternative}). Multiple benchmark DNNs in \cref{AllResults} were not implemented for MNIST due to its small resolution. As Stanford Dogs has 120 classes, we did not train the original ISNet on it; it would take about 2 months, while Faster ISNets required about 1 day (\cref{speed}). In Stanford Dogs, the synthetic bias also caused a strong shortcut learning tendency, demonstrated by the standard classifier's (ResNet50) strong performance drop upon bias removal or substitution by deceiving bias. Only the ISNets, GAIN (Grad-CAM optimization)\cite{GAIN}, and the segmentation-classification pipeline (U-Net+ResNet50) were robust to background bias and shortcut learning, being the only architectures whose performances remained effectively the same across the 3 test scenarios. However, GAIN was overshadowed by the pipeline's results. Conversely, some Faster ISNets could even surpass the pipeline, like the original ISNet did in \cite{ISNet}; thus, unlike GAIN, they did not trade accuracy for robustness. Instead of penalizing LRP, RRR and the ISNet Softmax Grad*Input ablation (\cref{Gradient ISNet}) minimize backgrounds in input gradients and Gradient*Input\cite{GradInput} heatmaps, respectively. With large images (e.g., 224x224) and deeper backbones (e.g., ResNet50), these 2 explanations become noisy\cite{ISNet}. Thus, the ISNet Softmax Grad*Input did not effectively converge in Stanford Dogs. Meanwhile, RRR minimized its less restrictive loss function\cite{RRR} by reducing its input gradients' magnitude, instead of promoting foreground (dogs) focus. The same phenomena were observed and justified in \cite{ISNet}. We chose biased Stanford Dogs to evaluate LDS due to the higher difficulty of its optimization task, considering strong background bias and 120 classes. Showing that LRP optimization is indeed more difficult in this dataset, on it, the Selective and Stochastic ISNets performances lagged behind the segmentation-classification pipeline. However, LDS, created to improve LRP loss convergence, allowed the models to match the pipeline. LDS had little effect on the Dual ISNet. It matched the pipeline without LDS, indicating that the DNN was already effectively converging, and that LDS was unnecessary for it.

\begin{figure}[!h]
\includegraphics[width=1\textwidth]{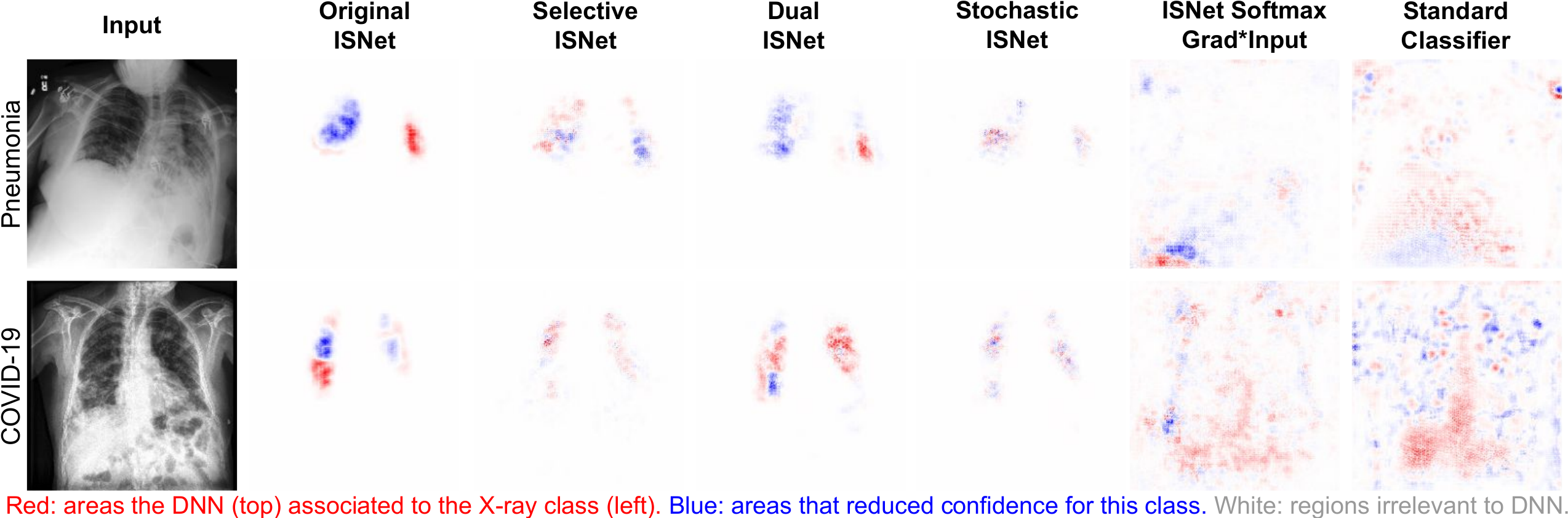}
\centering
\caption{LRP heatmaps for X-rays. The pneumonia X-ray has a clear background bias, a mark over the right shoulder. It did not influence the Faster or Original ISNets.}
\label{heatmapsFig1}
\end{figure}

Without synthetic bias, we trained DNNs to classify COVID-19, pneumonia and normal on the mixed X-ray dataset from \cite{ISNet}. It has an o.o.d. test partition (X-rays from external hospitals), and improvements in o.o.d. performance reflect increased background bias robustness and shortcut learning reduction\cite{ISNet,ShortcutLearning}. Here, foreground means lungs. The task's results in \cref{AllResults} are in line with past studies employing o.o.d. validation \cite{ISNet,bassi2021covid19,NatureCovidBias}. Conversely, i.i.d. COVID-19 detection performances can be deceitfully high and biased\cite{NatureCovidBias,ShortcutCovid}. We report maF1 as mean$\pm$std. \Cref{moreResults} shows additional performance metrics and \cref{statistical} explains our statistical analyses. We compared the Faster ISNet to the original ISNet and all benchmark DNNs trained in \cite{ISNet} for COVID-19 detection. The DenseNet121 was the backbone for ISNets and most benchmark DNNs\cite{ISNet}. \Cref{heatmapsFig1} presents heatmap examples, and \cref{heatmaps} analyses heatmaps for all tasks and DNNs, in light of the quantitative results in \cref{AllResults}. X-ray heatmaps revealed considerable background bias (e.g., text and marks) attention for all DNNs, except for the original ISNet, Faster ISNets, and segmentation-classification pipeline. Supporting the LRP maps, these DNNs had the highest o.o.d. F1-Scores in COVID-19 detection. Consistently, they were also robust to synthetic background bias. Although GAIN was robust to bias in Stanford Dogs, it suffered spurious mapping in COVID-19 classification, learning to produce deceiving GradCAM heatmaps that hid background attention to deceitfully reduce GAIN's losses\cite{ISNet}. Grad-CAM optimization may unpredictably lead to spurious mapping, but LRP optimization is immune to it\cite{ISNet}. All Faster ISNets performed comparably to the original ISNet, significantly surpassing the pipeline, whose decisions considered shortcuts like erased backgrounds and lung borders\cite{ISNet}. \Cref{speed} compared the ISNets' training speeds for different numbers of classes (c) in an application. Up to c=3, all ISNets have comparable speed. At c=4, Faster ISNets start being significantly faster than the original. At c=30, they are 10 to 20 times faster. 

\section{Conclusion}

This study introduced a simple, efficient and model-agnostic LRP implementation (LRP Flex), which explains arbitrary DNNs or converts them into ISNets; we improved the ISNet Loss and accelerated hyper-parameter tuning; we introduced LDS, improving LRP optimization convergence and accuracy; and we proposed the Faster ISNets. The original ISNet study elucidated the theoretical fundamentals of LRP optimization and justified why the ISNet surpasses multiple state-of-the-art benchmark DNNs, which include attention mechanisms and the optimization of Grad-CAM, input gradients and Gradient*Input\cite{ISNet}. Our results support their theoretical analyses. In both studies, the ISNets and the segmentation-classification pipeline were the only implemented DNNs that were consistently robust to background bias\cite{ISNet}. However, relying on 2 DNNs, the pipeline is much slower and heavier at inference; ISNets are about 2$\times$ faster and 5$\times$ smaller (with DenseNet121 backbone)\cite{ISNet}. They add no run-time computational cost. Matching the original ISNet, the Faster ISNet was robust to background bias, hindered shortcut learning, and consistently achieved the best o.o.d. generalization in all experiments. Unlike the original model, Faster ISNets' training time is independent of applications' number of classes. For 120 classes, they train about 50$\times$ faster than the ISNet; for 1000, they can reduce a 1-year training time to 1 day. By saving time, computational resources and energy, the Faster ISNet makes LRP optimization viable for a new plethora of applications.

\bibliography{sample}

\begin{thebibliography}{10}

\bibitem{ShortcutLearning}
R.~Geirhos, J.-H. Jacobsen, C.~Michaelis, R.~Zemel, W.~Brendel, M.~Bethge, and F.~Wichmann, ``Shortcut learning in deep neural networks,'' {\em Nature Machine Intelligence}, vol.~2, pp.~665--673, 11 2020.

\bibitem{NatureCovidBias}
A.~J. DeGrave, J.~D. Janizek, and S.-I. Lee, ``Ai for radiographic covid-19 detection selects shortcuts over signal,'' {\em Nat Mach Intell}, vol.~3, p.~610–619, 05/2021 2021.

\bibitem{ISNet}
P.~R. A.~S. Bassi, S.~S.~J. Dertkigil, and A.~Cavalli, ``Improving deep neural network generalization and robustness to background bias via layer-wise relevance propagation optimization,'' {\em Nature Communications}, vol.~15, Jan. 2024.

\bibitem{BrixiaSet}
A.~Signoroni, M.~Savardi, S.~Benini, N.~Adami, R.~Leonardi, P.~Gibellini, F.~Vaccher, M.~Ravanelli, A.~Borghesi, R.~Maroldi, and D.~Farina, ``Bs-net: learning covid-19 pneumonia severity on a large chest x-ray dataset,'' {\em Medical Image Analysis}, vol.~71, p.~102046, 03 2021.

\bibitem{GitCovidSet}
J.~P. Cohen, P.~Morrison, and L.~Dao, ``Covid-19 image data collection,'' {\em arXiv 2003.11597}, 2020.

\bibitem{reviewCovid}
A.~Shoeibi, M.~Khodatars, M.~Jafari, N.~Ghassemi, D.~Sadeghi, P.~Moridian, A.~Khadem, R.~Alizadehsani, S.~Hussain, A.~Zare, Z.~A. Sani, F.~Khozeimeh, S.~Nahavandi, U.~R. Acharya, and J.~M. Gorriz, ``Automated detection and forecasting of covid-19 using deep learning techniques: A review,'' {\em Neurocomputing}, vol.~577, p.~127317, 2024.

\bibitem{ShortcutCovid}
J.~López-Cabrera, J.~Portal~Diaz, R.~Orozco, O.~Lovelle, and M.~Perez-Diaz, ``Current limitations to identify covid‑19 using artificial intelligence with chest x‑ray imaging (part ii). the shortcut learning problem,'' {\em Health and Technology}, vol.~11, 10 2021.

\bibitem{critic}
G.~Maguolo and L.~Nanni, ``A critic evaluation of methods for covid-19 automatic detection from x-ray images,'' {\em Information Fusion}, vol.~76, pp.~1--7, 2021.

\bibitem{LRP}
S.~Bach, A.~Binder, G.~Montavon, F.~Klauschen, K.-R. Müller, and W.~Samek, ``On pixel-wise explanations for non-linear classifier decisions by layer-wise relevance propagation,'' {\em PLOS ONE}, vol.~10, pp.~1--46, 07 2015.

\bibitem{DenseNet}
G.~{Huang}, Z.~{Liu}, L.~{Van Der Maaten}, and K.~Q. {Weinberger}, ``Densely connected convolutional networks,'' in {\em 2017 IEEE Conference on Computer Vision and Pattern Recognition (CVPR)}, pp.~2261--2269, 2017.

\bibitem{ResNet}
K.~He, X.~Zhang, S.~Ren, and J.~Sun, ``Deep residual learning for image recognition,'' pp.~770--778, 06 2016.

\bibitem{mnist}
L.~Deng, ``The mnist database of handwritten digit images for machine learning research [best of the web],'' {\em IEEE Signal Processing Magazine}, vol.~29, no.~6, pp.~141--142, 2012.

\bibitem{StanfordDogs}
A.~Khosla, N.~Jayadevaprakash, B.~Yao, and L.~Fei-Fei, ``Novel dataset for fine-grained image categorization,'' in {\em First Workshop on Fine-Grained Visual Categorization, IEEE Conference on Computer Vision and Pattern Recognition}, (Colorado Springs, CO), June 2011.

\bibitem{LRPBook}
G.~Montavon, A.~Binder, S.~Lapuschkin, W.~Samek, and K.-R. M{\"u}ller, ``Layer-wise relevance propagation: An overview,'' in {\em Explainable AI: Interpreting, Explaining and Visualizing Deep Learning}, pp.~193--209, Springer International Publishing, 2019.

\bibitem{LRPZb}
G.~Montavon, S.~Lapuschkin, A.~Binder, W.~Samek, and K.-R. Müller, ``Explaining nonlinear classification decisions with deep taylor decomposition,'' {\em Pattern Recognition}, vol.~65, p.~211–222, 05 2017.

\bibitem{unet}
O.~Ronneberger, P.~Fischer, and T.~Brox, ``U-net: Convolutional networks for biomedical image segmentation,'' in {\em Medical Image Computing and Computer-Assisted Intervention -- MICCAI 2015} (N.~Navab, J.~Hornegger, W.~M. Wells, and A.~F. Frangi, eds.), (Cham), pp.~234--241, Springer International Publishing, 2015.

\bibitem{deepMAC}
V.~Birodkar, Z.~Lu, S.~Li, V.~Rathod, and J.~Huang, ``The surprising impact of mask-head architecture on novel class segmentation,'' in {\em 2021 IEEE/CVF International Conference on Computer Vision (ICCV)}, IEEE, Oct. 2021.

\bibitem{VGG}
K.~Simonyan and A.~Zisserman, ``Very deep convolutional networks for large-scale image recognition,'' in {\em International Conference on Learning Representations}, 2015.

\bibitem{GWRP}
S.~Qiu, ``Global weighted average pooling bridges pixel-level localization and image-level classification,'' 2018.

\bibitem{RRR}
A.~S. Ross, M.~C. Hughes, and F.~Doshi-Velez, ``Right for the right reasons: Training differentiable models by constraining their explanations,'' in {\em Proceedings of the Twenty-Sixth International Joint Conference on Artificial Intelligence, {IJCAI-17}}, pp.~2662--2670, 2017.

\bibitem{MulticlassAUC}
D.~J. Hand and R.~J. Till, ``A simple generalisation of the area under the roc curve for multiple class classification problems,'' {\em Mach. Learn.}, vol.~45, p.~171–186, Oct. 2001.

\bibitem{VisionTransformer}
A.~Dosovitskiy, L.~Beyer, A.~Kolesnikov, D.~Weissenborn, X.~Zhai, T.~Unterthiner, M.~Dehghani, M.~Minderer, G.~Heigold, S.~Gelly, J.~Uszkoreit, and N.~Houlsby, ``An image is worth 16x16 words: Transformers for image recognition at scale,'' in {\em International Conference on Learning Representations}, 2021.

\bibitem{AGNet}
J.~Schlemper, O.~Oktay, M.~Schaap, M.~Heinrich, B.~Kainz, B.~Glocker, and D.~Rueckert, ``Attention gated networks: Learning to leverage salient regions in medical images,'' 2018.

\bibitem{GAIN}
K.~Li, Z.~Wu, K.-C. Peng, J.~Ernst, and Y.~Fu, ``Tell me where to look: Guided attention inference network,'' 03 2018.

\bibitem{GradInput}
A.~Shrikumar, P.~Greenside, A.~Shcherbina, and A.~Kundaje, ``Not just a black box: Learning important features through propagating activation differences,'' {\em ArXiv}, vol.~abs/1605.01713, 2016.

\bibitem{bassi2021covid19}
P.~R. A.~S. Bassi and R.~Attux, ``Covid-19 detection using chest x-rays: is lung segmentation important for generalization?,'' {\em Research on Biomedical Engineering}, vol.~38, p.~1121–1139, Nov. 2022.

\bibitem{welford}
B.~P. Welford, ``Note on a method for calculating corrected sums of squares and products,'' {\em Technometrics}, vol.~4, no.~3, pp.~419--420, 1962.

\bibitem{MultiTaskOriginal}
R.~Caruana, ``Multitask learning,'' {\em Machine Learning}, vol.~28, 07 1997.

\bibitem{attentionSurvey}
M.-H. Guo, T.-X. Xu, J.-J. Liu, Z.-N. Liu, P.-T. Jiang, T.-J. Mu, S.-H. Zhang, R.~Martin, M.-M. Cheng, and S.-M. Hu, ``Attention mechanisms in computer vision: A survey,'' {\em Computational Visual Media}, 03 2022.

\bibitem{MultiTask1}
Z.~Kong, M.~He, Q.~Luo, X.~Huang, P.~Wei, Y.~Cheng, L.~Chen, Y.~Liang, Y.~Lu, X.~Li, and J.~Chen, ``Multi-task classification and segmentation for explicable capsule endoscopy diagnostics,'' {\em Frontiers in Molecular Biosciences}, vol.~8, 2021.

\bibitem{MultiTask2}
Y.~Zhou, H.~Chen, Y.~Li, Q.~Liu, X.~Xu, S.~Wang, and P.-T. Yap, ``Multi-task learning for segmentation and classification of tumors in 3d automated breast ultrasound images,'' {\em Medical Image Analysis}, vol.~70, p.~101918, 11 2020.

\bibitem{Grad-CAM}
R.~R. Selvaraju, M.~Cogswell, A.~Das, R.~Vedantam, D.~Parikh, and D.~Batra, ``Grad-cam: Visual explanations from deep networks via gradient-based localization,'' in {\em 2017 IEEE International Conference on Computer Vision (ICCV)}, pp.~618--626, 2017.

\bibitem{bayesianEstimator}
D.~Zhang, J.~Wang, and X.~Zhao, ``Estimating the uncertainty of average f1 scores,'' {\em Proceedings of the 2015 International Conference on The Theory of Information Retrieval}, 2015.

\bibitem{pymc}
J.~Salvatier, T.~Wiecki, and C.~Fonnesbeck, ``Probabilistic programming in python using pymc3,'' {\em PeerJ Computer Science}, 2016.

\bibitem{NUTS}
M.~D. Homan and A.~Gelman, ``The no-u-turn sampler: Adaptively setting path lengths in hamiltonian monte carlo,'' {\em J. Mach. Learn. Res.}, vol.~15, p.~1593–1623, Jan. 2014.

\bibitem{imagenet}
J.~Deng, W.~Dong, R.~Socher, L.~Li, K.~Li, and L.~Fei-Fei, ``Imagenet: A large-scale hierarchical image database,'' in {\em 2009 IEEE Computer Society Conference on Computer Vision and Pattern Recognition Workshops (CVPR Workshops)}, (Los Alamitos, CA, USA), pp.~248--255, IEEE Computer Society, jun 2009.

\bibitem{FirstPaperCovid}
P.~R. A.~S. Bassi and R.~Attux, ``A deep convolutional neural network for covid-19 detection using chest x-rays,'' {\em Research on Biomedical Engineering}, Apr 2021.

\bibitem{irvin2019chexpert}
J.~Irvin, P.~Rajpurkar, M.~Ko, Y.~Yu, S.~Ciurea-Ilcus, C.~Chute, H.~Marklund, B.~Haghgoo, R.~Ball, K.~Shpanskaya, J.~Seekins, D.~Mong, S.~Halabi, J.~Sandberg, R.~Jones, D.~Larson, C.~Langlotz, B.~Patel, M.~Lungren, and A.~Ng, ``Chexpert: A large chest radiograph dataset with uncertainty labels and expert comparison,'' {\em Proceedings of the AAAI Conference on Artificial Intelligence}, vol.~33, pp.~590--597, 07 2019.

\bibitem{BimcvSet}
M.~de~la Iglesia~Vayá, J.~M. Saborit, J.~A. Montell, A.~Pertusa, A.~Bustos, M.~Cazorla, J.~Galant, X.~Barber, D.~Orozco-Beltrán, F.~García-García, M.~Caparrós, G.~González, and J.~M. Salinas, ``Bimcv covid-19+: a large annotated dataset of rx and ct images from covid-19 patients,'' 2020.

\bibitem{chex14}
X.~Wang, Y.~Peng, L.~Lu, Z.~Lu, M.~Bagheri, and R.~M. Summers, ``Chestx-ray8: Hospital-scale chest x-ray database and benchmarks on weakly-supervised classification and localization of common thorax diseases,'' {\em 2017 IEEE Conference on Computer Vision and Pattern Recognition (CVPR)}, Jul 2017.

\bibitem{ChineseDataset1}
S.~Jaeger, S.~Candemir, S.~Antani, Y.-X.~J. W{\'a}ng, P.-X. Lu, and G.~R. Thoma, ``Two public chest x-ray datasets for computer-aided screening of pulmonary diseases.,'' {\em Quantitative imaging in medicine and surgery}, vol.~4 6, pp.~475--7, 2014.

\bibitem{DeLongAUC}
E.~R. DeLong, D.~M. DeLong, and D.~L. Clarke-Pearson, ``Comparing the areas under two or more correlated receiver operating characteristic curves: a nonparametric approach.,'' {\em Biometrics}, vol.~44 3, pp.~837--45, 1988.

\end{thebibliography}

\appendix
\section{LRP-Flex}
\label{FlexAppendix}

The original ISNet study reformulated the LRP-$\varepsilon$ relevance propagation procedure (Supplementary Note 8.2 of\cite{ISNet}) to make the denoising property of $\varepsilon$ more explicit. The reformulation objective was to better justify the empirically observed advantage of LRP-$\varepsilon$ optimization over the optimization of LRP-0 (Gradient*Input), an explanation methodology that is noisy for deep networks\cite{LRPBook}. First, it defined the tensor $\mathbf{G^{L}}$ (with elements $G^{L}_{j}$) as the LRP-$\varepsilon$ relevance at the input of the DNN layer L ($\mathbf{R^{L}}$, with elements $R^{L}_{j}$), divided (element-wise) by the layer's input ($\mathbf{a^{L}}$, with elements $a^{L}_{j}$). This definition is shown in \cref{RG}, considering all layers in the DNN (from 0, the first layer, to Lmax, the output layer). Accordingly, the LRP-$\varepsilon$ heatmap ($\mathbf{R^{0}}$, the LRP relevance at the network's input) is given by the element-wise multiplication ($\odot$) between $\mathbf{G^{0}}$ and the DNN input, $\mathbf{a^{0}}$ (\cref{R0}).

\begin{gather}
\label{RG}
G^{L}_{j}=R^{L}_{j}/a^{L}_{j} \mbox{ $\forall$ } L \in [0,1,...,\textrm{Lmax}]\\
\label{R0}
\mathbf{R^{0}}=\mathbf{G^{0}} \odot \mathbf{a^{0}}
\end{gather}

For a neural network utilizing only ReLU nonlinearities, we can find a simple recursive equation to backpropagate $\mathbf{G^{L}}$\cite{ISNet}. We rearrange the LRP-$\varepsilon$ rule (\cref{LRP-eSuper}) considering \cref{RG}, resulting in \cref{deduction}. Here, $w_{jk}^{L}$ represents the layer L weight connecting its j-th input to its k-th output; $w_{0k}^{L}$ indicates the k-th bias parameters of layer L (considering $a_{0}^{L}=1$); $z_{k}^{L}$ is the k-th output of layer L, before its ReLU activation; and $a^{L+1}_{k}$ is the k-th input of layer L+1, which matches $\mathrm{ReLU}(z_{k}^{L})$.

\begin{gather}
\label{LRP-eSuper}
R_{j}^{L}=\sum_{k}\frac{w_{jk}^{L}a_{j}^{L}}{z_{k}^{L}+\textrm{sign}(z_{k}^{L})\epsilon}R_{k}^{L+1} \\
\label{deduction}
\begin{multlined}
G_{j}^{L}=R^{L}_{j}/a^{L}_{j}=
\sum_{k}\frac{w_{jk}^{L}a_{k}^{L+1}G_{k}^{L+1}}{z_{k}^{L}+\mathrm{sign}(z_{k}^{L})\varepsilon}=
\sum_{k}w_{jk}^{L}\frac{\mathrm{ReLU}(z_{k}^{L})}{z_{k}^{L}+\mathrm{sign}(z_{k}^{L})\varepsilon}G_{k}^{L+1}
\end{multlined}
\end{gather}

\Cref{AS} defines the Attenuated Step function ($A(\cdot)$)\cite{ISNet}. Accordingly, we reduce \cref{deduction} into \cref{GBack}, which is a simple rule to backpropagate $\mathbf{G^{L}}$ through the DNN layers. We can recursively apply \cref{GBack} until the network's input layer, producing $\mathbf{G^{0}}$. Afterward, considering \cref{R0}, an element-wise multiplication of $\mathbf{G^{0}}$ and the DNN input ($\mathbf{a^{0}}$) generates the LRP-$\varepsilon$ heatmap. When explaining a DNN's logit ($z_{c}^{\textrm{Lmax}}$), the chosen logit's relevance is set to the logit value itself ($z_{c}^{\textrm{Lmax}}$), while the other logits' relevances are set to zero\cite{LRP}. As we are explaining a logit, the last DNN layer is seen as linear during the relevance propagation, and LRP-0 is a standard choice for such layers\cite{LRPBook}. Thus, the LRP relevance at the layer's inputs ($a^{\textrm{Lmax}}_{j}$) becomes $R^{\textrm{Lmax}}_{j}=w_{jc}^{\textrm{Lmax}}a_{j}^{\textrm{Lmax}}$. Accordingly and following \cref{RG}, $\mathbf{G^{\textrm{Lmax}}}$ at the input of the DNN last layer matches the layers' weights (\cref{GInitial}).

\begin{gather}
\label{GBack}
G_{j}^{L}=\sum_{k}w_{jk}^{L}A(z_{k}^{L})G_{k}^{L+1} \\
\label{AS}
\mbox{where: } A(z_{k}^{L})=\frac{\mathrm{ReLU}(z_{k}^{L})}{z_{k}^{L}+\mathrm{sign}(z_{k}^{L})\varepsilon}=
\begin{cases}
\frac{z_{k}^{L}}{z_{k}^{L}+\varepsilon}, \mbox{ if } z_{k}^{L} \geq 0 \\
0,  \mbox{ otherwise }
\end{cases}\\
\label{GInitial}
G_{j}^{\textrm{Lmax}}=w_{jc}^{\textrm{Lmax}}
\end{gather}

\cref{GBack} is similar to the gradient backpropagation rule for fully-connected layers with ReLU activations, shown in \cref{gradRule} (where $z_{c}^{\textrm{Lmax}}$ indicates a logit whose gradient is being back-propagated). Indeed, the only difference between the two recursive equations is the substitution of the Attenuated Step function in \cref{GBack} by the Unit Step function ($H(\cdot)$, defined in \cref{unitStep}). When $\varepsilon=0$, the Attenuated Step function matches the Unit Step, and the two equations become equivalent. Moreover, the initial values for the gradient and for G match (Equations \ref{GInitial} and \ref{gradInitial}). Therefore, if $\varepsilon=0$, the quantity G is the gradient ($G_{j}^{L}=\frac{\partial z_{c}^{\textrm{Lmax}}}{\partial a_{j}^{L}} \mbox{ $\forall$ } L \in [0,1,...,\textrm{Lmax}]$). Accordingly, it is known that LRP-0 heatmaps match the element-wise multiplication between the DNN input ($\mathbf{a^{0}}$) and the input gradient ($\nabla_{\mathbf{a^{0}}}z_{c}^{\textrm{Lmax}}$)\cite{LRPBook}. But this equivalence does not hold if $\varepsilon \neq 0$, i.e., when LRP-$\varepsilon$ substitutes LRP-0\cite{ISNet}.

\begin{gather}
\label{gradRule}
\frac{\partial z_{c}^{\textrm{Lmax}}}{\partial a_{j}^{L}}= \sum_{k} w_{jk}^{L} H(z_{k}^{L}) \frac{\partial z_{c}^{\textrm{Lmax}}}{\partial a_{k}^{L+1}}\\
\mbox{where: } H(z_{k}^{L})=
\label{unitStep}
\begin{cases}
1, \mbox{ if } z_{k}^{L} > 0 \\
0, \mbox{ if } z_{k}^{L} < 0 
\end{cases} \\
\label{gradInitial}
\frac{\partial z_{c}^{\textrm{Lmax}}}{\partial a_{j}^{\textrm{Lmax}}}=w_{jc}^{\textrm{Lmax}}
\end{gather}

The original ISNet paper proposed the alternative LRP-$\varepsilon$ formulation (defined by Equations \ref{GInitial}, \ref{GBack} and \ref{R0}) to justify the empirically observed advantage of LRP-$\varepsilon$ optimization (ISNet) over optimizing LRP-0 or Gradient*Input (ISNet Grad*Input). LRP-0 and Gradient*Input are noisy for deep neural networks, and LRP-$\varepsilon$ produces less noisy explanations, which are easier to optimize. Accordingly, the ISNet displayed a more stable and easier loss convergence with respect to the ISNet Grad*Input, especially with very deep classifiers. Thus, the ISNet surpassed the ISNet Grad*Input in terms of accuracy and resistance to background bias\cite{ISNet}. The Attenuated Step (\cref{AS}) is a monotonically increasing function, which approximates the Unit Step for negative or high arguments. However, unlike the Unit Step, it reduces the influence of small neuron activations (small $\mathrm{ReLU}(z_{k}^{L})$) on the back-propagated quantity ($G_{j}^{L}$) and on the final heatmap. The higher the $\varepsilon$, the stronger the attenuation\cite{ISNet}. Therefore, $\varepsilon$ controls a denoising process, justified by the Deep Taylor Decomposition framework, where LRP-$\varepsilon$ represents a reduction in the first-order Taylor approximation error in relation to LRP-0 and Gradient*Input\cite{LRPBook}. The process attenuates the explanations' noise and improves their coherence and contextualization\cite{ISNet,LRPBook}.

In this study, we show that the alternative LRP formulation defined in Equations \ref{GInitial}, \ref{GBack} and \ref{R0} (conceived as a theoretical justification in the original ISNet paper\cite{ISNet}) can derive an alternative LRP implementation, LRP-Flex. The implementation is exceedingly simple and model agnostic, being readily applicable to any deep neural network with only ReLU nonlinearities in its hidden layers.

First, the gradient backpropagation rule in \cref{gradRule} can be separated into two parts: a gradient back-propagation through the ReLU activation (\cref{reluBack}), followed by the back-propagation of $\frac{\partial z_{c}^{\textrm{Lmax}}}{\partial z_{k}^{L}}$ through the fully-connected layer L (\cref{FCBack}).

\begin{gather}
\label{reluBack}
\frac{\partial z_{c}^{\textrm{Lmax}}}{\partial z_{k}^{L}}= H(z_{k}^{L}) \frac{\partial z_{c}^{\textrm{Lmax}}}{\partial a_{k}^{L+1}}\\
\label{FCBack}
\frac{\partial z_{c}^{\textrm{Lmax}}}{\partial a_{j}^{L}}= \sum_{k} w_{jk}^{L} \frac{\partial z_{c}^{\textrm{Lmax}}}{\partial z_{k}^{L}}
\end{gather}

Similarly, we can separate the alternative LRP-$\varepsilon$ G backpropagation rule (\cref{GBack}) in two sequential steps: backpropagation of G through the ReLU function (\cref{reluBackG}), and through the preceding fully-connected layer (\ref{FCBackG}).

\begin{gather}
\label{reluBackG}
U_{k}^{L} = A(z_{k}^{L}) G_{k}^{L+1} \\
\label{FCBackG}
G_{j}^{L}= \sum_{k} w_{jk}^{L} U_{k}^{L}
\end{gather}

Equations \ref{FCBackG} and \ref{FCBack} show that the backpropagation procedure through the fully-connected layer is the same for both the gradient and G. Meanwhile, the backpropagation step through the ReLU activation is not. \Cref{reluBackG} represents a modification in the standard backward pass through ReLU functions (\cref{reluBack}): instead of multiplying the back-propagated quantity ($G_{k}^{L+1}$) by the Unit Step $H(z_{k}^{L})$, we multiply it (element-wise) by the Attenuated Step $A(z_{k}^{L})$. Therefore, with a simple modification of the ReLU's backward step (substituting $H(\cdot)$ by $A(\cdot)$), automatic gradient backpropagation engines can backpropagate G instead of the actual gradient. Notice that the Attenuated Step can be directly calculated from the output of the ReLU (\cref{AltAS}). In summary, we can easily leverage the automatic differentiation capabilities offered by deep learning libraries to produce LRP heatmaps. Gradient backpropagation algorithms tend to be fast, highly optimized, and readily applicable to new DNN architectures. To obtain G at the DNN input ($\mathbf{G^{0}}$), we modify the ReLU back-propagation rule, and request the backpropagation engine to compute the gradient of the DNN's explained logit ($z_{c}^{\textrm{Lmax}}$) with respect to the network's input ($\mathbf{a^{0}}$). Then, the LRP-$\varepsilon$ heatmap is obtained as the element-wise multiplication between $\mathbf{G^{0}}$ and the DNN input (\cref{R0}).

\begin{gather}
\label{AltAS}
A(z_{k}^{L})=A(ReLU(z_{k}^{L}))=\frac{\mathrm{ReLU}(z_{k}^{L})}{\mathrm{ReLU}(z_{k}^{L})+\varepsilon}
\end{gather}

In PyTorch, a backward hook (applied with register\_full\_backward\_hook) can modify the ReLU module's backward pass, implementing \cref{reluBackG} and making the automatic differentiation procedure backpropagate G instead of the gradient. To calculate the Attenuated Step function, we can use the output of the ReLU ($ReLU(z_{k}^{L})$) as argument (\cref{AltAS}). This output can be stored by a forward hook (applied with register\_forward\_hook) during the forward pass. The backward hooks activate automatically in backward passes, but they can be deactivated (e.g., by changing a global flag variable) during a standard loss backpropagation in the training procedure. With a recursive function, we can easily apply the hooks to all ReLU layers in an instantiated network. We must only avoid in-place operations inside the DNN from changing the ReLU outputs stored during the forward pass. Moreover, a single ReLU layer must not be re-utilized multiple times in the network. Finally, the backpropagation algorithm in PyTorch is differentiable. Therefore, LRP-Flex can be used to quickly and easily explain a trained DNN's decisions, and to produce the differentiable LRP heatmaps employed for training the original, Selective and Stochastic ISNets.

\cref{GInitial} is valid for creating a heatmap that explains a DNN's logit, such as the explanations in the original and Stochastic ISNets. However, the Selective ISNet maps explain a softmax-based quantity ($\eta_{c}$, Main Article \cref{nFrac}) instead of a logit ($z_{c}^{\textrm{Lmax}}$). Main Article \cref{selectiveEquation1} explains the calculation of the relevance at the input of the DNN's last layer ($R^{\textrm{Lmax}}_{j}$) when explaining $\eta_{c}$. Here, we demonstrate that, when explaining $\eta_{c}$, the quantity $G^{\textrm{Lmax}}_{j}$, defined as $R^{\textrm{Lmax}}_{j}/a^{\textrm{Lmax}}_{j}$ (\cref{RG}), is equivalent to the gradient of $\eta_{c}$ with respect to the inputs of the last DNN layer (\cref{gradEta}). 

\begin{equation}
\label{gradEta}
G^{\textrm{Lmax}}_{j}=\frac{\partial \eta_{c}}{\partial a_{j}^{\textrm{Lmax}}}
\end{equation}

To prove this equality, we first take the derivative of $\eta_{c}$ with respect to $z_{c,c'}$ (defined in Main Article \cref{selectiveEquation2}), as shown in \cref{nFracDeriv}. Afterward, we can take derivative of $\eta_{c}$ with respect to the inputs of the last DNN layer ($a_{j}^{\textrm{Lmax}}$), using the calculus chain rule, as displayed in \cref{nFracDeriv2}.

\begin{gather}
\label{nFracDeriv}
\frac{\partial \eta_{c}}{\partial z_{c,c'}}=\frac{\partial [- \ln (\sum_{c''\neq c} e^{-z_{c,c''}})]}{\partial z_{c,c'}} =\frac{e^{-z_{c,c'}}}{\sum_{c''\neq c} e^{-z_{c,c''}}}\\
\begin{multlined}
\label{nFracDeriv2}
\frac{\partial \eta_{c}}{\partial a_{j}^{\textrm{Lmax}}}=\sum_{c'}\frac{\partial \eta_{c}}{\partial z_{c,c'}} \frac{\partial z_{c,c'}}{\partial a_{j}^{\textrm{Lmax}}}=
\sum_{c'} \frac{e^{-z_{c,c'}}}{\sum_{c''\neq c} e^{-z_{c,c''}}} (w_{jc}^{\textrm{Lmax}}-w_{jc'}^{\textrm{Lmax}})
\end{multlined}
\end{gather}

Now, we combine Main Article Equations \ref{selectiveEquation1} and \ref{selectiveEquation3} into \cref{rearrangedR}. Main Article Equations \ref{selectiveEquation1} and \ref{selectiveEquation3} are the standard rules to propagate LRP relevance until $a_{j}^{\textrm{Lmax}}$ when explaining $\eta_{c}$\cite{LRPBook}. Here, we set the stabilizers $\mu$ and $\epsilon$ to zero in Main Article Equations \ref{selectiveEquation1} and \ref{selectiveEquation3}. Notice that $\mu$ is not part of the standard propagation rule (it was added to improve numerical stability in this work), and setting $\epsilon=0$ would represent the use of LRP-0 in the last DNN layer, a standard choice\cite{LRPBook}. Thus, we remove both terms, resulting in \cref{rearrangedR2}.

\begin{gather}
\label{rearrangedR}
R_{j}^{\textrm{Lmax}}=\sum_{c'}\frac{(w_{jc}^{\textrm{Lmax}}-w_{jc'}^{\textrm{Lmax}})a_{j}^{\textrm{Lmax}}}{z_{c,c'}+\textrm{sign}(z_{c,c'})\epsilon} \frac{z_{c,c'} e^{-z_{c,c'}}}{\sum_{c''\neq c}e^{-z_{c,c''}}+\mu}\\
\label{rearrangedR2}
R_{j}^{\textrm{Lmax}}=a_{j}^{\textrm{Lmax}} \sum_{c'}\frac{ e^{-z_{c,c'}}}{\sum_{c''\neq c}e^{-z_{c,c''}}}(w_{jc}^{\textrm{Lmax}}-w_{jc'}^{\textrm{Lmax}})
\end{gather}

To derive $G_{j}^{\textrm{Lmax}}$ from \cref{rearrangedR} (which represents standard LRP relevance propagation rules\cite{LRPBook}), we divide both of its sides by $a_{j}^{\textrm{Lmax}}$, producing \cref{almost} (considering the definition $R_{j}^{L}=a_{j}^{L}G_{j}^{L}$ in \cref{RG}). Finally, comparing Equations \ref{nFracDeriv2} and \ref{almost}, we prove the equality $G^{\textrm{Lmax}}_{j}=\frac{\partial \eta_{c}}{\partial a_{j}^{\textrm{Lmax}}}$ shown in \cref{gradEta}.

\begin{gather}
\label{almost}
G^{\textrm{Lmax}}_{j}=R_{j}^{\textrm{Lmax}}/a_{j}^{\textrm{Lmax}}=\sum_{c'}\frac{ e^{-z_{c,c'}}}{\sum_{c''\neq c}e^{-z_{c,c''}}}(w_{jc}^{\textrm{Lmax}}-w_{jc'}^{\textrm{Lmax}})=\frac{\partial \eta_{c}}{\partial a_{j}^{\textrm{Lmax}}}
\end{gather}

Exploiting \cref{gradEta}, to explain $\eta_{c}$ instead of $z_{c}^{\textrm{Lmax}}$, LRP-Flex could simply request the backpropagation engine for the gradient of $\eta_{c}$ with respect to the DNN input, applying the modified ReLU backpropagation rule in \cref{reluBackG}. Then, the resulting tensor ($\mathbf{G^{0}}$) is element-wise multiplied with the DNN's input, producing the LRP-$\varepsilon$ heatmap (\cref{R0}). However, $\eta_{c}$ backpropagation can be numerically unstable, as $\eta_{c}$ explodes when the softmax probability of class c approaches 0 or 1 (Main Article \cref{nFrac}). Therefore, to ensure numerical stability, we backpropagate a bounded version of $\eta_{c}$, defined in \cref{boundEta}. In the equation, $P_{c}$ is the probability of class c (i.e., the Softmax output for logit $z_{c}^{\textrm{Lmax}}$, Main Article \cref{softmax}), and $\mu$ is a positive constant. The higher the $\mu$ 
hyper-parameter, the smaller the maximum magnitude of $\tilde{\eta_{c}}$ (and of its gradient). We empirically set $\mu=0.01$.

\begin{gather}
\label{boundEta}
\tilde{\eta_{c}}=ln(P_{c}+\mu)-\ln(1-P_{c}+\mu)
\end{gather}

For simplicity, the equations in this section considered fully-connected layers. However, convolutions can be represented as fully connected layers. Moreover, other common layers, such as batch normalization, dropout and pooling can also be expressed as fully connected layers or fused with adjacent fully-connected layers/convolutions\cite{ISNet}. Thus, the reasoning in this section and the LRP-Flex implementation are not restricted to fully-connected DNNs. Indeed, LRP-Flex just requires the DNN's hidden layers to contain only ReLU nonlinearities. However, LRP-$\varepsilon$ was designed for DNNs with ReLU activations anyway\cite{LRP}, and such architectures compose a substantial proportion of the state-of-the-art (e.g., ResNets and DenseNets).

To use LRP-z$^{B}$ for the first DNN layer, we stop the LRP-Flex modified backward pass at the output of the first DNN layer ($\mathbf{a^{1}}$, the output of the network's first ReLU). In other words, we change Step 3 of the LRP-Flex procedure (Main Article \cref{Flex}): we now request the backpropagation engine for the gradient of the DNN logit (or of $\tilde{\eta_{c}}$) with respect to $\mathbf{a^{1}}$, employing the modified ReLU backpropagation rule in Step 1. Step 4 is also changed: we element-wise multiply the tensor obtained in the modified Step 3, $\mathbf{G^{1}}$, with $\mathbf{a^{1}}$, obtaining the LRP-$\varepsilon$ relevance at the output of the first DNN layer ($\mathbf{R^{1}}=\mathbf{G^{1}}\odot \mathbf{a^{1}}$). Finally, $\mathbf{R^{1}}$ can be back-propagated through the first DNN layer according to the standard LRP-z$^{B}$ rule\cite{LRPBook} (e.g., following the implementation in the LRP Block\cite{ISNet}), generating the LRP heatmap ($\mathbf{R^{0}}$). The use of LRP-z$^{B}$ is not mandatory for training an ISNet \cite{ISNet}, nor is it necessary for the Faster ISNet.

\section{LRP Block Improvements}
\label{lrpBlock}

The minimization of the heatmap loss requires gradient backpropagation through the LRP rules. The paper presenting the ISNet\cite{ISNet} introduced the LRP Block, an efficient structure that performs LRP over a classifier, and produces differentiable heatmaps. The differentiability allows popular deep learning frameworks (e.g., PyTorch) to perform automatic backpropagation of the heatmap loss gradient. Please refer to the study introducing the ISNet for a detailed explanation of the Block\cite{ISNet}. It presents computationally efficient procedures to propagate relevance through standard DNN layers, namely: fully-connected, convolutional, max and average pooling, dropout, and batch normalization. All procedures are equivalent to LRP-$\varepsilon$. For a convolutional or fully-connected layer with ReLU activation, the strategy is described below (extracted from\cite{ISNet}). The LRP Block also has a LRP-z$^{B}$ implementation, for the first DNN layer\cite{ISNet}.

\begin{enumerate}
    \item Forward pass the layer L input, $\mathbf{x}_{L}$, through a copy of layer L, generating its output tensor $\mathbf{z}$ (without activation). Use parameter sharing, use the detach() function on the biases. Get $\mathbf{x}_{L}$ via a skip connection with layer L.
    \item For LRP-$\epsilon$, modify each $\mathbf{z}$ tensor element ($z$) by adding $\textrm{sign}(z)\epsilon$ to it. Defining the layer L output relevance as $\mathbf{R}_{L}$, perform its element-wise division by $\mathbf{z}$: $\mathbf{s}=\mathbf{R}_{L}/\mathbf{z}$.
    \item Forward pass the quantity $\mathbf{s}$ through a transposed version of the layer L (linear layer with transposed weights or transposed convolution), generating the tensor $\mathbf{c}$. Use parameter sharing but set biases to 0.
    \item Obtain the relevance at the input of layer L by performing an element-wise product between $\mathbf{c}$ and the layer input values, $\mathbf{x}_{L}$ (i.e., the output of layer L-1): $\mathbf{R}_{L-1}=\mathbf{x}_{L} \odot \mathbf{c}$.
\end{enumerate}

The function detach() is a PyTorch function, in the procedure it makes the bias parameter in the layer L copy not affect the layer L bias gradient. Notice that, in the procedure, multiple skip connections tie the classifier and LRP Block, as LRP propagation rules consider the inputs and/or outputs of all classifier layers. The LRP Block has no exclusive trainable parameter, because the LRP rules only employ the classifier's parameters. As heatmap generation is not necessary after training, the structure can be deactivated or deleted at run-time\cite{ISNet}.

In this study, we introduced an alternative to the LRP Block, LRP-Flex, a simpler and model agnostic implementation of LRP. However, we also present improvements to the LRP Block, which allow its utilization to implement the Dual and Selective ISNets (the Stochastic ISNet requires no specific modification to the original LRP Block). Notice that, as LRP-Flex does not compute LRP-z$^{+}$ heatmaps, we exclusively use the LRP Block for implementing the Dual ISNet. In summary, the improvements we present are: (1) the efficient implementation of LRP-z$^{+}$ propagation rules, required by the Dual ISNet; (2) inclusion of the capacity to explain the softmax-related quantity $\eta_{c}$ (Main Article \cref{SelectiveISNet}), creating the more class-selective LRP heatmaps required by the Selective ISNet; (3) implementation of the LRP Block for the ResNet architecture.

\subsection{Dual ISNet LRP Block}

The Dual ISNet optimizes a LRP-$\epsilon$ joint heatmap and a LRP-z$^{+}$ joint map (Main Article \cref{DualISNet}). The LRP-$\epsilon$ joint heatmap can be produced with the standard LRP Block\cite{ISNet}, by just setting the relevance at the DNN output to the logits' values (i.e., $R_{k}^{\textrm{Lmax}+1}=z_{k}^{\textrm{Lmax}} \forall k \in [1,2,...,C]$). I.e., for the output layer, set $\mathbf{R}_{L}=\mathbf{z}$ in the 4-step procedure from \cref{lrpBlock}. Meanwhile, to produce the LRP-$z^{+}$ joint heatmap, we implemented a LRP Block based on the LRP-$z^{+}$ rules. For numerical stability during training, we utilized a stabilizer constant ($\epsilon$) in LRP-$z^{+}$ (Main Article \cref{lrp-z+_equation}). We set it to $10^{-2}$, the same value we used in LRP-$\epsilon$. 

Most layers in a standard DNN have only positive inputs, due to a preceding ReLU activation. In this case, we can simplify the quantity $(w_{jk}^{L}a_{j}^{L})^{+}$ in Main Article \cref{lrp-z+_equation} to $(w_{jk}^{L})^{+}a_{j}^{L}$. Therefore, to implement the LRP-$z^{+}$ procedure for a fully-connected/convolutional layer L we must simply alter the four-step sequence in \cref{lrpBlock} to consider a new L$^{+}$ layer, instead of L (see the procedure below). All layer L positive parameters (weights and biases) are shared with layer L$^{+}$, while its negative parameters are set to 0 in the new layer. Notice that this LRP-$z^{+}$ propagation procedure has essentially the same computational cost as the original LRP-$\epsilon$ procedure.

\begin{enumerate}
    \item Forward pass the layer L input, $\mathbf{x}_{L}$, through layer L$^{+}$, generating the quantity $\mathbf{z}_{pos}$ (without activation). Use parameter sharing, use the detach() function on the biases. Get $\mathbf{x}_{L}$ via a skip connection with layer L.
    \item Modify each $\mathbf{z}_{pos}$ tensor element by adding $\epsilon$ to it. Defining the layer L output relevance as $\mathbf{R}_{L}$, perform its element-wise division by $\mathbf{z}_{pos}$: $\mathbf{s}_{pos}=\mathbf{R}_{L}/\mathbf{z}_{pos}$.
    \item Forward pass the quantity $\mathbf{s}_{pos}$ through a transposed version of the layer L$^{+}$ (linear layer with transposed weights or transposed convolution), generating the tensor $\mathbf{c}_{pos}$. Use parameter sharing but set biases to 0.
    \item Obtain the relevance at the input of layer L by performing an element-wise product between $\mathbf{c}_{pos}$ and the layer input values, $\mathbf{x}_{L}$ (i.e., the output of layer L-1): $\mathbf{R}_{L-1}=\mathbf{x}_{L} \odot \mathbf{c}_{pos}$.
\end{enumerate}

When a layer receives negative inputs, the aforementioned LRP-$z^{+}$ implementation is not adequate. It would produce negative relevance, which is not acceptable for LRP-$z^{+}$. Therefore, a more complex procedure is adopted when we verify negative inputs. We describe it below for a fully-connected/convolutional layer L, noticing that it uses two forward passes and two transpose passes, twice as the original four-step procedure for LRP-$\varepsilon$. By $\mathbf{X}^{+}$ we indicate a tensor that shares all positive elements with tensor $\mathbf{X}$ and is zero where $\mathbf{X}$ has negative values. Conversely, $\mathbf{X}^{-}$ only shares the negative elements in $\mathbf{X}$ and is 0 otherwise. With L$^{+}$/L$^{-}$ we represent layers that share all positive/negative parameters with layer L, and present zeros where L has negative/positive parameters. 

\begin{enumerate}
    \item The layer L input is $\mathbf{x}_{L}$. Forward pass $\mathbf{x}_{L}^{+}$ through layer L$^{+}$, and $\mathbf{x}_{L}^{-}$ through layer L$^{-}$. Sum the two results, producing $\mathbf{z}_{pos}$, the denominator in Main Article \cref{lrp-z+_equation} before the addition of $\epsilon$. Use parameter sharing, utilize the detach() function on the biases, do not employ non-linear activations. Get $\mathbf{x}_{L}$ via a skip connection with layer L.
    \item Modify each $\mathbf{z}_{pos}$ tensor element by adding $\epsilon$ to it. Defining the layer L output relevance as $\mathbf{R}_{L}$, perform its element-wise division by the modified $\mathbf{z}_{pos}$ tensor, producing the new tensor $\mathbf{s}_{pos}$: $\mathbf{s}_{pos}=\mathbf{R}_{L}/\mathbf{z}_{pos}$.
    \item Forward pass the quantity $\mathbf{s}_{pos}$ through a transposed version of the layer L$^{+}$ (linear layer with transposed weights or transposed convolution), generating the tensor $\mathbf{c_{pos}}$. Moreover, forward pass $\mathbf{s}_{pos}$ through a transposed version of the layer L$^{-}$, producing the tensor $\mathbf{c_{neg}}$. Use parameter sharing but set all biases to 0.
    \item Obtain the relevance at the input of layer L by performing an element-wise product between $\mathbf{c_{pos}}$ and $\mathbf{x}_{L}^{+}$ and summing it to the element-wise product between $\mathbf{c_{neg}}$ and $\mathbf{x}_{L}^{-}$. I.e., $\mathbf{R}_{L-1}=\mathbf{x}_{L}^{+} \odot \mathbf{c}_{pos} + \mathbf{x}_{L}^{-} \odot \mathbf{c}_{neg}$.
\end{enumerate}

For the creation of the LRP-$z^{+}$ heatmap, we still use the LRP-$z^{B}$ rule for the input layer, as is done for the LRP-$\epsilon$ map. The implementation of the rule is described in the original ISNet study\cite{ISNet}. If the relevance signal it receives only has positive values, LRP-$z^{B}$ cannot produce negative relevance, making it adequate for producing the LRP-$z^{+}$ heatmap. The same is true for the LRP Block procedure for MaxPool layers\cite{ISNet}. Meanwhile, batch normalization operations can be fused with adjacent convolutional layers, producing an equivalent convolutional layer. Moreover, average pooling can also be reformulated as an equivalent convolution. The LRP Block LRP-$\epsilon$ relevance propagation procedures for such layers\cite{ISNet} consider their equivalent convolutions and employ the four-step procedure described in \cref{lrpBlock}. Therefore, to implement LRP-$z^{+}$ for batch normalization and pooling we can simply utilize the procedures previously described in this subsection, considering the layer L as the equivalent convolution. Please refer to the original ISNet paper\cite{ISNet} for details on the implementation of the equivalent convolution. Finally, the two procedures in this section can also be applied to the Dual ISNet output layer, by just setting DNN output relevance, $\mathbf{R}_{L}$, equal to $\mathbf{z}_{pos}$.

\subsection{Selective ISNet LRP Block}

To explain $\eta_{c}$ instead of the logit $z_{c}^{\textrm{Lmax}}$, we use Main Article \cref{selectiveEquation1} when propagating the LRP relevance through the last classifier layer (Lmax). Notice that, to obtain $R_{j}^{\textrm{Lmax}}$ (Main Article \cref{selectiveEquation1}), we must first calculate $z_{c,c'}$ (Main Article \cref{selectiveEquation2}), then $R_{c,c'}$ (Main Article \cref{selectiveEquation3}). Accordingly, for the last DNN layer, we substitute the 4-step procedure in \cref{lrpBlock} by the procedure below.

\begin{enumerate}
    \item Create a new layer, define its weights and biases according to Main Article \cref{selectiveEquation2}. I.e., repeat the weights ($w_{jc}^{\textrm{Lmax}}$) and bias ($w_{0c}^{\textrm{Lmax}}$) for the lowest logit's neuron C (number of classes) times, and subtract all the original output layer's weights (weights$=w_{jc}^{\textrm{Lmax}}-w_{jc'}^{\textrm{Lmax}}$) and biases (biases$=w_{0c}^{\textrm{Lmax}}-w_{0c'}^{\textrm{Lmax}}$). Use parameter sharing between the new and original layers.
    \item Forward pass the last layer inputs, $a_{j}^{\textrm{Lmax}}$, through the new layer, generating its outputs $z_{c,c'}$ (without activation). Get $a_{j}^{\textrm{Lmax}}$ via a skip connection with the original last DNN layer.
    \item Obtain the output relevance for the new layer ($R_{c,c'}$) from $z_{c,c'}$, according to Main Article \cref{selectiveEquation3}.
    \item To implement Main Article \cref{selectiveEquation1}, utilize the standard LRP Block 4-step procedure for linear layers (\cref{lrpBlock}) to propagate the $R_{c,c'}$ relevance through the new layer, whose parameters were calculated in Step 1. Consider $R_{c,c'}$, $a_{j}^{\textrm{Lmax}}$, and $z_{c,c'}$ as the new layer's output relevance, inputs, and outputs, respectively. The 4-step procedure will calculate its input relevance, $R_{j}^{\textrm{Lmax}}$.
\end{enumerate}

\subsection{ResNet-based ISNet}

Layer-Wise Relevance Propagation is an efficient technique, which can explain virtually any deep neural network architecture\cite{LRP}. Accordingly, the ISNet's background relevance minimization procedure is applicable to diverse classifier architectures. The original ISNet paper\cite{ISNet} utilized DenseNets\cite{DenseNet}, due to their depth and computational efficiency. It also implemented ISNets based on the classical and popular VGG backbone\cite{VGG}. Finally, it proposed implementations for arbitrary feed-forward networks without skip connections, composed of common sequences of standard neural network layers (including ReLU, fully-connected, convolutional, batch normalization, dropout, and pooling)\cite{ISNet}.

Residual neural networks (ResNets)\cite{ResNet} and their derivatives are among the most popular architectures for very deep neural networks. The models utilize skip connections to improve signal and gradient propagation, allowing the effective optimization of networks with more than 100 layers\cite{ResNet}. A ResNet requires more memory and training time than a DenseNet with a similar number of layers. Therefore, the improved training efficiency introduced by the Faster ISNet is exceedingly beneficial for ResNet-based ISNets. We implemented the LRP Block for all standard ResNet versions, namely ResNet18, 34, 50, 101 and 152. It can be utilized to implement the original ISNet or any of the Faster ISNet variants. The considered backbones are based on PyTorch's official code for ResNets. 

\begin{figure}[!h]
\includegraphics[width=0.5\textwidth]{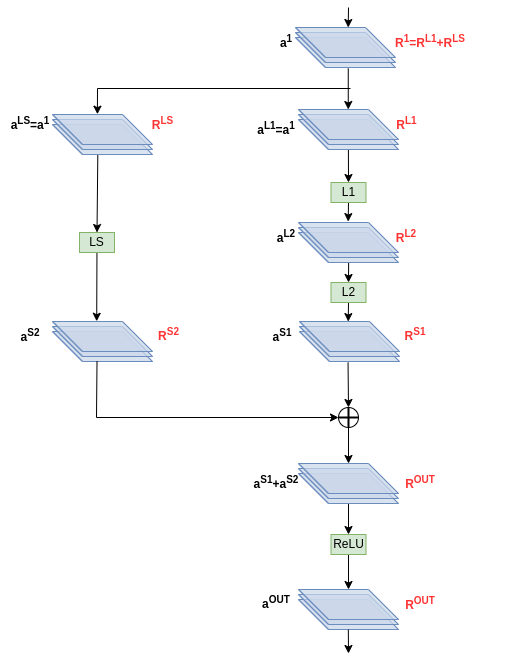}
\centering
\caption{Illustration of residual block. Neural network layers are indicated in green, black text indicates the signal, while red text shows the corresponding LRP relevance.}
\label{resnet}
\end{figure}

Most of the layer sequences inside a ResNet are already present in a DenseNet or VGG. Thus, their corresponding LRP Block procedures were already presented by the original ISNet study\cite{ISNet}. The exception is the ResNet's summation junction, at the end of each skip connection. A ResNet is mainly composed of a sequence of residual blocks, which are small sequences of convolutional layers in parallel with a shortcut connection. The connection ties the block's input and output, where an element-wise summation is used to connect the two paths. The shortcut path can be defined by an identity operation, or by a convolutional layer with 1x1 sized kernels, which is used for down-sampling\cite{ResNet}. The convolutions have ReLU activations, which can be preceded by batch normalization. \Cref{resnet} shows a basic residual block illustration, which follows PyTorch's implementation. L1, L2 and LS represent three sequences of layers. In ResNet18, L1 means convolution followed by batch normalization and ReLU, while L2 indicates convolution followed by batch normalization. The shortcut path, LS, can be an identity function or a sequence of convolution and batch normalization. The LRP Block procedure to propagate relevance through the layers contained in L1, L2 and LS are detailed in the original ISNet paper\cite{ISNet}.

Following the standard LRP procedure for shortcut connections\cite{ISNet,LRPBook}, the residual block input LRP relevance is the sum of the relevances obtained at the inputs of the principal and the shortcut path ($\mathbf{R}^{1}=\mathbf{R}^{LS}+\mathbf{R}^{L1}$). Moreover, the LRP relevance at the output of the summation junction ($\mathbf{R}^{OUT}=[R_{j}^{OUT}]$) is proportionally distributed for the two junction's inputs\cite{LRPBook} ($\mathbf{R}^{S1}=[R_{j}^{S1}]$, and $\mathbf{R}^{S2}=[R_{j}^{S2}]$). Consider that the summation junction sums $\mathbf{a}^{S1}+\mathbf{a}^{S2}=[a_{j}^{S1}+a_{j}^{S2}]$. Then, for LRP-$\epsilon$ we have:

\begin{gather}
    R_{j}^{S1}=\frac{a_{j}^{S1}}{a_{j}^{S1}+a_{j}^{S2}+\epsilon \textrm{ sign}(a_{j}^{S1}+a_{j}^{S2})}R_{j}^{OUT}\\
    R_{j}^{S2}=\frac{a_{j}^{S2}}{a_{j}^{S1}+a_{j}^{S2}+\epsilon \textrm{ sign}(a_{j}^{S1}+a_{j}^{S2})}R_{j}^{OUT}
\end{gather}

For LRP-$z^{+}$ (Dual ISNet) we utilize:
\begin{gather}
    R_{j}^{S1}=\frac{\textrm{max}(a_{j}^{S1},0)}{\textrm{max}(a_{j}^{S1},0)+\textrm{max}(a_{j}^{S2},0)+\epsilon}R_{j}^{OUT}\\
    R_{j}^{S2}=\frac{\textrm{max}(a_{j}^{S2},0)}{\textrm{max}(a_{j}^{S1},0)+\textrm{max}(a_{j}^{S2},0)+\epsilon}R_{j}^{OUT}
\end{gather}

The relevance propagation through the bottleneck residual block (used in ResNets with 50 layers or more)\cite{ResNet} is very similar to the procedure for the standard residual block. The bottleneck block has three convolutional layers in its main path, instead of two. However, all layers in the path represent standard sequences involving convolution, batch normalization and ReLU. Propagation procedures for such sequences are detailed in the original ISNet study\cite{ISNet}.

\section{ISNet Loss and Hyper-Parameter Selection}
\label{heatmapLoss}

The ISNet loss, $L_{IS}$, is a linear combination of a standard classification loss (e.g., cross-entropy), $L_{C}$, and the heatmap loss, $L_{LRP}$\cite{ISNet}. The latter quantifies the amount of undesired background attention in LRP heatmaps, created by the LRP Block. The most important hyper-parameter for training the ISNet is P, which balances the influence of $L_{C}$ and $L_{LRP}$. The DNN is sensible to this parameter, low values allow attention to background bias, while high P can reduce training speed. The original ISNet study presents strategies to tune P, considering access or no access to o.o.d. validation data\cite{ISNet}.

\begin{equation}
L_{IS}=(1-P).L_{C}+P.L_{LRP}, \mbox{ where } 0 \le P \le 1
\end{equation}

The heatmap loss is composed of two terms, the heatmap background loss, $L_{1}$, and the heatmap foreground loss, $L_{2}$, as shown in \cref{HLoss}. The first term is responsible for measuring background attention, while the second avoids a zero solution to $L_{LRP}$ (when the neural network produces heatmaps valued zero everywhere), or exploding heatmap relevance. The two balancing hyper-parameters, $w_{1}$ and $w_{2}$, have little influence over the ISNet behavior. Thus, a fine search is not necessary to define them. We set both to 1 in the MNIST experiments, and we set $w_{1}=1$ and $w_{2}=3$ in the other applications, matching the values in the original ISNet\cite{ISNet}. However, in preliminary tests, we found similar results when setting both parameters to 1 in all applications.

\begin{equation}
\label{HLoss}
L_{LRP}=w_{1}.L_{1}+w_{2}.L_{2}, \mbox{ where } 0 < w_{1} \mbox{ and } 0 < w_{2}
\end{equation}

The calculation of the heatmap background loss involves a few processing steps, summarized below. For a detailed mathematical definition and justification for each procedure, please refer to the paper presenting the original ISNet\cite{ISNet}.

\begin{enumerate}
    \item Absolute heatmaps: take the absolute value of the LRP heatmaps. 
    \item Normalized absolute heatmaps: normalize the absolute heatmaps, dividing them by the absolute average relevance in their foreground.
    \item Segmented heatmaps: in the normalized absolute heatmaps, set all foreground relevance to zero, by element-wise multiplying them by inverted segmentation targets (i.e., figures valued one in the background and 0 in the foreground).
    \item Raw background attention scores: use Global Weighted Ranked Pooling\cite{GWRP} (GWRP) over the segmented heatmaps, obtaining one scalar score per map channel.
    \item Activated scores: pass the raw scores through the non-linear function $f(x)=x/(x+E)$, where E is a constant hyper-parameter, normally set as 1.
    \item Background attention loss: calculate the cross-entropy between the activated scores and zero, i.e., apply the function $g(x)=-ln(1-x)$
    \item $L_{1}$: calculate the average background attention loss for all heatmaps in the training mini-batch.
\end{enumerate}

GWRP\cite{GWRP} (step 4 in the procedure) is a weighted arithmetic mean (in the two spatial dimensions) of the elements in the segmented heatmaps (output of step 3). GWRP ranks every element in the heatmaps in descending order. Following this order, the elements' weights in the summation decay exponentially, according to a hyper-parameter $d$, valued between 0 and 1. Thus, high relevance elements in the LRP heatmap background highly increment the heatmap loss. If $d=0$, GWRP is equivalent to max pooling. When it is set as 1, GWRP is the same as average pooling. The original ISNet study suggested increasing $d$ for improving training stability and reducing it to increase resistance to background bias (especially avoiding small high-attention regions in the background). The study also showed procedures to tune the $d$ hyper-parameter, with or without access to o.o.d. validation data\cite{ISNet}.

Finally, the second term of the heatmap loss, the foreground loss, $L_{2}$, is calculated with the procedure summarized below. Again, a detailed mathematical explanation is provided in the paper presenting the ISNet\cite{ISNet}. 

\begin{enumerate}
    \item Absolute heatmaps: take absolute value of the LRP heatmaps. 
    \item Segmented maps: element-wise multiply the absolute heatmaps by the segmentation targets, which are valued 1 in the foreground and 0 in the background.
    \item Absolute foreground relevance: sum all elements in the segmented maps, obtaining the total (absolute) foreground relevance per-heatmap.
    \item Square losses: if the absolute foreground relevance, $s$, is smaller than a constant hyper-parameter, $C_{1}$, the correspondent square loss is $(C_{1}-s)^{2}/C_{1}^{2}$. If it is larger than $C_{2}$ ($C_{2}>C_{1}>0$), its correspondent square loss is $(C_{2}-s)^{2}/C_{1}^{2}$. If $C_{1}>s>C_{2}$, the correspondent loss is 0. 
    \item $L_{2}$: take the average of all square losses, considering all heatmaps in the mini-batch.
\end{enumerate}

The $L_{2}$ loss is valued 0 when the heatmap absolute foreground relevance is in the range $[C_{1},C_{2}]$, and it raises quadratically when it exits the range. The loss guarantees that the heatmap relevances will stay within a natural range, avoiding undesirable solutions to the background heatmap loss ($L_{1}$), such as solutions involving heatmaps valued zero or exploding heatmap elements. 

\subsection{ISNet Loss Modification}

\begin{figure}[!h]
\includegraphics[width=0.5\textwidth]{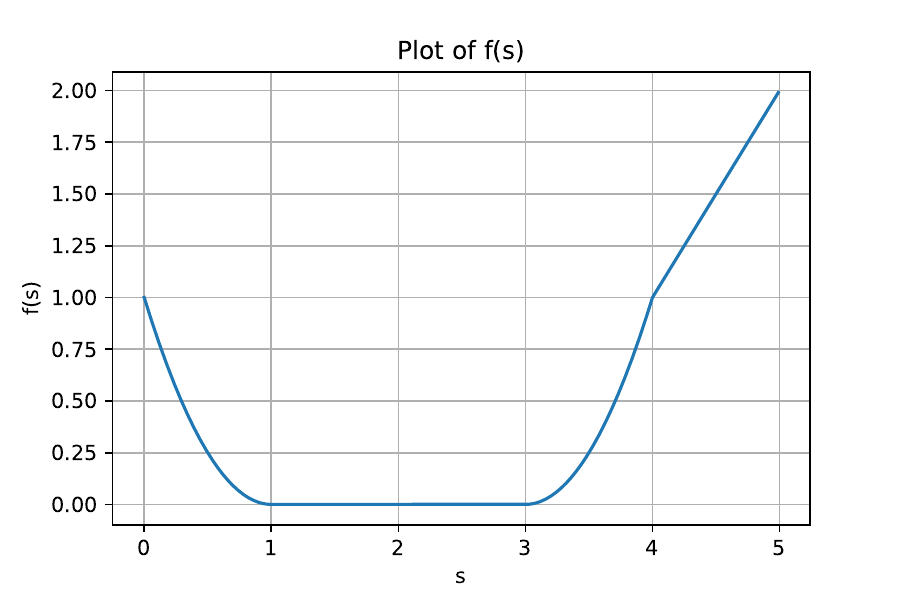}
\centering
\caption{Plot of the f(s) function, for $C_{1}=1$ and $C_{2}=3$.}
\label{lossPlot}
\end{figure}

Here, we propose a modification to the original ISNet loss, specifically, to $L_{2}$. We employed it in all ISNets trained in this study. The gradient of square losses in Step 4 of the procedure above can become exceedingly high if the absolute foreground relevance considerably exceeds the expected natural relevance range ($s>>C_{2}$). This situation can happen in the beginning of the training procedure, especially if $C_{2}$ is small (e.g., due to weight decay, as explained in \cref{tuneCut}). Therefore, for training stability, we change the loss function in Step 4, switching from a square to a linear loss for high values of $s$. \Cref{newLoss} expresses the new loss function in Step 4, and \cref{lossPlot} plots it.

\begin{equation}
\label{newLoss}
    f(s)=\begin{cases}
(C_{1}-s)^{2}/C_{1}^{2} \mbox{ if } s<C_{1} \\
0 \mbox{ if } C_{1}\le s\le C_{2} \\
(C_{2}-s)^{2}/C_{1}^{2}  \mbox{ if } C_{2}\le s\le C_{2}+C_{1} \\
1+s-(C_{2}+C_{1})  \mbox{ if } C_{2}+C_{1}\le s
    \end{cases}
\end{equation}

\subsection{Heatmap Range Selection}
\label{tuneCut}

The $[C_{1},C_{2}]$ range objective is to approximate the natural interval of the total absolute relevance (sum of the absolute values of all heatmap elements). Therefore, it is derived from the analysis of heatmaps from a standard classifier (non-ISNet), which follows the ISNet backbone architecture, and shares its training dataset\cite{ISNet}. This natural range is similar for the same classifier architecture and heatmap creation procedure, even for diverse training datasets and classification tasks\cite{ISNet}. However, the ideal $[C_{1},C_{2}]$ can drastically change when we alter the ISNet explanation strategy (e.g., changing from the original ISNet to the Selective ISNet or ISNet Softmax Grad*Input), or the ISNet classifier backbone (e.g., from DenseNet to ResNet or VGG).

To accelerate hyper-parameter search, we have empirically derived a heuristic to define the hyper-parameters $C_{1}$ and $C_{2}$. First, train a standard classifier (non-ISNet) with the same architecture and training algorithm as the ISNet classifier, using the same training dataset, but for a small number of epochs (we used 4). Since weight decay can reduce the natural range of LRP relevance through training, we suggest training the non-ISNet network for longer when using the technique (up to the same number of epochs that will be used for the ISNet). Afterward, train the standard classifier for one epoch more, but generate its heatmaps for all training samples (using the same heatmap creation procedure that the ISNet will employ). For each heatmap, calculate the total absolute relevance ($\sum_{j}abs(H_{j})$, where $H_{j}$ is a heatmap element). During this last epoch, use Welford's online algorithm\cite{welford} to estimate the mean and standard deviation of the total absolute relevance across all training samples. The online method reduces memory consumption. Finally, define $C_{1}$ and $C_{2}$ according to the equations below, where M indicates the calculated mean, S the standard deviation, min the minimum function, and max the maximum function.

\begin{gather}
\label{c1}
C_{1}=\textrm{max}(M/5,M-3S)\\
\label{c2}
C_{2}=\textrm{min}(25C_{1},M+3S)
\end{gather}

The hyper-parameters are defined according to absolute total relevance in a non-ISNet network, because we desire to understand what the natural range of values in a heatmap are. Since we are not interested in the standard classifier's ability to concentrate attention on the region of interest, we do not derive $[C_{1},C_{2}]$ from the network's foreground absolute relevance. Instead, the ISNet optimization process will concentrate heatmap relevance inside the foreground, making the ISNet's foreground absolute relevance range approximate the standard classifier's total absolute relevance range.

Ideally, the Equations \ref{c1} and \ref{c2} select the range $[M-3S,M+3S]$, which should capture a substantial portion of the total absolute relevance we observed for the standard classifier during training. However, we must use the maximum operation in \cref{c1} to avoid $C_{1}$ becoming negative or too close to zero. Moreover, the minimum function in \cref{c2} avoids an exceedingly major difference between the two hyper-parameters, which could allow unnatural solutions for the heatmap loss\cite{ISNet}. The quantities $M/5$ and $25C_{1}$ were empirically defined during preliminary tests.

The natural range of relevance values can change across DNN layers, due to LRP relevance absorption\cite{LRPBook}. Therefore, if we penalize LRP heatmaps at the inputs of multiple hidden layers (LRP Deep Supervision or LDS), we must define a $C_{1}$ $C_{2}$ pair for each of these layers. To do so, instead of only monitoring the standard classifier's input-level heatmaps, we calculate the mean (M) and standard deviation (S) of the total absolute LRP relevance independently (and in parallel) for all supervised layers. Accordingly, Equations \ref{c1} and \ref{c2} can provide the parameters $[C_{1},C_{2}]$ for each penalized layer.

Instabilities in gradient backpropagation (e.g., vanishing and exploding gradients) can be harmful for LRP relevance propagation. If the model's input gradients are problematic around the data point (the input image), first-order Taylor expansions considering nearby reference points will probably also have issues. LRP is based on the approximation of a series of local Taylor expansions\cite{LRPBook}. Thus, LRP heatmaps may also explode or vanish when gradients explode or vanish. In such cases, Equations \ref{c1} and \ref{c2} will return extremely high or low values, which are not adequate for training an ISNet. However, this problem can be solved by standard techniques that avoid vanishing and exploding gradients, such as adequate weight initialization and batch normalization.

\section{Benchmark DNNs}
\label{alternative}

For both the COVID-19 detection and the biased MNIST classification tasks, we compare the Faster ISNet formulations to standard classifiers, which employ the same architecture as the ISNet classifier (ResNet18 for MNIST and DenseNet121 for COVID-19 detection). The comparison will allow us to assess the effect of the ISNet's LRP optimization on shortcut learning, generalization capacity and attention to background bias. Moreover, we also compare the new models to the original ISNet\cite{ISNet} in all applications except for dog breed classification, where the original model's training time would be excessive. The low image resolution of MNIST is not detrimental to input gradients and Gradient*Input explanations. Therefore, in the application we also compare the Faster ISNet to the ISNet Softmax Grad*Input (\cref{Gradient ISNet}) and the Right for the Right Reasons neural network\cite{RRR} (RRR). Like the ISNet, both utilize the ResNet18 backbone in MNIST, DenseNet121 in COVID-19 detection and ResNet50 in Stanford Dogs. RRR is based on the optimization of input gradients during training. It uses a square loss to minimize the background region of the gradients. This region is defined according to segmentation masks\cite{RRR}. Here, input gradients are the gradients of the summed log probabilities for all possible classes, with respect to the DNN input\cite{RRR}. As the ISNet, RRR's objective is to produce classifiers whose decisions ignore background features. 

This study utilizes the same COVID-19 detection dataset used in the original ISNet paper\cite{ISNet}. Therefore, we can directly compare the Faster ISNet performance to multiple state-of-the-art neural network architectures that control classifier attention, which were already tested in the original ISNet report\cite{ISNet}. They are: the original ISNet\cite{ISNet}, a standard segmentation-classification pipeline\cite{ISNet}, multi-task DNNs\cite{MultiTaskOriginal}, attention-gated neural networks (AG-Sononet)\cite{AGNet}, Guided Attention Inference Networks (GAIN)\cite{GAIN}, RRR\cite{RRR}, and vision transformer (ViT-B/16)\cite{VisionTransformer}. We implemented the same benchmark models in the Stanford Dogs dataset. Because MNIST images and the corresponding deep feature maps are exceedingly small, most of the baseline models were not adequate for MNIST classification. However, although we do not train the segmentation-classification pipeline on MNIST, we train and test a classifier (ResNet18) on the unbiased (standard) MNIST dataset, simulating the classifier performance on perfectly segmented images. More details about implementations, hyper-parameters, and training procedures for the alternative deep neural networks on the COVID-19 classification dataset can be found in the original ISNet paper\cite{ISNet}. Meanwhile, their training procedure in Stanford Dogs is explained in \cref{TrainingProcedure}. Below, we summarize these different state-of-the-art methodologies\cite{ISNet}.

First, we compare the ISNet to a standard technique used for datasets containing background bias, the segmentation-classification pipeline: a deep semantic segmenter identifies the image foreground (e.g., lungs), the segmenter output is thresholded and used to substitute the image's background by zeros, and finally a deep classifier classifies the segmented image. In the ISNet original study, this segmentation-classification pipeline was the only alternative model that could also be insensitive to the presence of background bias\cite{ISNet}. However, the network was still surpassed by the ISNet in all experiments in the study\cite{ISNet}. The pipeline represents a much larger and slower model during run-time, due to the utilization of two deep neural networks. The segmenter in the model is a standard U-Net\cite{unet}, and the classifier a DenseNet121\cite{DenseNet} in COVID-19 detection and a ResNet50 in Stanford Dogs, matching the ISNet classifier. The U-Net is a semantic segmenter especially designed for biomedical images\cite{unet}, composed of a contracting signal path, followed by an expanding path. The paths are connected by multiple skip connections. The U-Net was pretrained on the COVID-19 dataset, using its segmentation targets\cite{ISNet}. Afterward, the segmenter parameters were frozen, and the classifier was trained with the same database.

The ISNet's background relevance minimization can be seen as a spatial attention mechanism (explanation-based) because it represents a technique designed to dynamically control a classifier's attention over the input space\cite{ISNet}. Accordingly, we compare the model to a state-of-the-art attention mechanism, the AG-Sononet\cite{AGNet}. Attention gated networks were designed to substitute the aforementioned pipeline of segmentation-classification, especially for biomedical images\cite{AGNet}. Its objective is to increase classifier focus on important image features and reduce attention to background clutter (features uncorrelated to the sample class). The AG-Sononet introduced attention gates, a structure that performs the element-wise multiplication of a convolutional layer's feature maps and attention coefficients. The coefficients are defined as a function of two tensors: the gated layer's feature map, and the output of a late convolutional layer (e.g., the last). The AG-Sononet is based on a VGG-16 classifier, and the considered implementation\cite{ISNet} followed the AG-Sononet official code\cite{AGNet}. The model exemplifies the paradigm of not using segmentation targets to guide the classifier attention. Like the AG-Sononet, most spatial attention mechanisms rely solely on the classification labels and the classification loss to teach the DNN what features are relevant in an image\cite{attentionSurvey}. The AG-Sononet architecture presents minimal computational overhead in relation to standard classifiers, and it was successfully used to avoid classifier focus on background clutter. The model was evaluated on fetal ultrasounds, which represent noisy and hard to interpret images\cite{AGNet}.

The Vision Transformer\cite{VisionTransformer} is a popular neural network architecture, built upon an attention mechanism: the transformer's self-attention. Like the AG-Sononet, it does not learn from segmentation targets. The Transformer was highly successful in natural language processing (NLP). Its attention mechanism relates diverse positions in an input sequence, generating a new representation of the sequence. Specifically, the network implemented here and in \cite{ISNet} is the base model ViT-B/16\cite{VisionTransformer}, because, with respect to alternative variants, its number of parameters is closer to the other benchmark networks\cite{ISNet}.

As an additional benchmark, we consider a multi-task DNN simultaneously performing semantic segmentation and classification\cite{MultiTaskOriginal}. Such networks begin as a shared sequence of convolutional layers, which fork into two independent paths, the classification head and the segmentation head. The first outputs classification scores, while the second produces semantic segmentation outputs. To optimize the entire network, the model utilizes a linear combination of a classification loss (employing classification targets) and a segmentation loss (using segmentation masks). The objective of the segmentation loss is simply the creation of semantic segmentation outputs. Thus, it is vastly different from the ISNet's heatmap loss, whose objective is the control of what input features influence the classification scores. Moreover, the segmentation outputs in the multi-task network are produced by an independent structure, the segmentation head. They are not explanations of the classifier scores, nor necessarily related to the classifier's behavior, like the LRP heatmaps optimized by the ISNet\cite{ISNet}.

Nevertheless, the original ISNet work\cite{ISNet} tested if the multi-task network can be resistant to background bias, because studies suggested that the segmentation task can help the classifier ignore irrelevant background clutter, improving accuracy\cite{MultiTask1,MultiTask2}. Examples of noisy and cluttered images, where multi-task learning improved classification, were capsule endoscopy pictures\cite{MultiTask1} and breast ultrasound\cite{MultiTask2}. The multi-task DNN implemented in the original ISNet study\cite{ISNet} is based on the U-Net architecture, with an additional classification head attached to the network bottleneck feature representation (end of the U-Net contracting path). The U-Net contracting path is based on the VGG-16 architecture. Thus, the classification head follows the VGG-16 last layers\cite{VGG}, consisting in average pooling (outputting 7x7 sized feature maps), followed by three dense layers, the first two with 4096 neurons. The classification head utilizes 50\% dropout layers after the first two dense layers, and it included batch normalization before each ReLU activation in the U-Net body to improve training convergence\cite{ISNet}.

Finally, the last benchmark DNN in COVID-19 detection and dog breed classification is the Guided Attention Inference Network (GAIN)\cite{GAIN}. The model was proposed to create a classifier whose Grad-CAM heatmaps more completely indicate the objects in the images, serving as priors for weakly supervised semantic segmentation\cite{GAIN}. As an additional benefit, GAIN's authors claim that the architecture can reduce classifier attention to background bias\cite{GAIN}. Like the ISNet, GAIN directly optimizes explanation heatmaps. However, it penalizes Grad-CAM heatmaps instead of LRP maps. Refer to the original ISNet study\cite{ISNet} for a theoretical comparison between the two methodologies, and for an analysis of the LRP optimization advantages. Besides the standard classification loss (e.g., cross-entropy), two loss functions penalize the Grad-CAM explanations, the attention mining loss and the external supervision loss\cite{Grad-CAM}. The network classifies an image, produces its Grad-CAM heatmaps for the ground-truth classes, and utilizes the explanations to remove the high attention region from the input image. When multiple classes are present, multiple heatmaps and altered images are generated. The new figures are classified again, and the attention mining loss is defined as the mean value of new ground-truth classes' logits\cite{Grad-CAM}. The loss minimization prompts the neural network to pay attention to all input image regions that are correlated to the sample labels, extending the Grad-CAM attention. The external supervision loss is the mean squared error between the Grad-CAM maps for the ground-truth classes and the respective segmentation targets. It is presented as an optional loss, which is important for resistance to background attention\cite{GAIN}. Thus, to maximize bias resistance, the GAIN implementation in the original ISNet paper\cite{ISNet} utilizes the external supervision loss for all training images. GAIN with the external supervision loss can also be called extended GAIN\cite{GAIN}. The classifier chosen as backbone for GAIN was the DenseNet121 in COVID-19 detection and the ResNet50 in dog breed classification, matching the ISNet classifier\cite{GAIN}. GAIN's Grad-CAM considers the last convolutional feature map in the DNN, the standard choice for GAIN\cite{GAIN}. Latter heatmaps contain higher-level semantics\cite{Grad-CAM}, and allow the utilization of more DNN layers (and flexibility) to satisfy the GAIN losses.

\subsection{ISNet Softmax Grad*Input}
\label{Gradient ISNet}

Gradient*Input explanations consist in the element-wise multiplication of the DNN input and the gradient of the logit, with respect to the input\cite{GradInput}. Originally, the explanation technique was proposed as a way to improve sharpness in comparison to input gradients\cite{GradInput}. In deep neural networks with only ReLU activations (except for the output layer), Gradient*Input is equivalent to LRP-0\cite{ISNet}. The equivalence was formally demonstrated, assuming numerical stability in the LRP-0 relevance propagation (in practice, the technique is unstable)\cite{ISNet}. LRP-$\varepsilon$ produces less noisy explanations of deep neural networks, when compared to Gradient*Input or LRP-0. Moreover, LRP-$\varepsilon$ explanations are more contextualized and coherent. Such advantages are justified by the deep Taylor decomposition (DTD) framework, which explains a neural network's output with a series of approximate first-order Taylor expansions\cite{ISNet}. From the DTD standpoint, in comparison to LRP-0 or Gradient*Input, LRP-$\varepsilon$ represents a reduction in the Taylor approximation error\cite{LRPBook,ISNet}. In the original ISNet study\cite{ISNet}, the ISNet was compared to the ISNet Grad*Input, an ablation model where the ISNet's LRP-$\varepsilon$ heatmaps were substituted by Gradient*Input explanations. Demonstrating the advantages of LRP-$\varepsilon$ optimization, the ISNet significantly surpassed the ISNet Grad*Input, producing models that were more resistant to background bias and more accurate, especially when deep architectures and large input images were considered\cite{ISNet}.

Like the original ISNet, the ISNet Grad*Input produced one heatmap per classifier output neuron, causing its training time to increase linearly with the number of classes in the classification task. Each heatmap corresponded to the input multiplied (element-wise) by the gradient of one logit. Here, the Selective ISNet uses LRP to explain a softmax-related quantity, $\eta_{c}$, avoiding the need to propagate multiple heatmaps per image (Main Article \cref{SelectiveISNet}). Accordingly, to avoid performing a gradient backpropagation for each logit, the ISNet Softmax Grad*Input explains the classifier Softmax output for only the winning class (c). I.e., the heatmaps are defined as the gradient of the Softmax score for c ($P_{c}$) with respect to the input ($\mathbf{X}$), element-wise multiplied with the input itself: $\mathbf{X} \odot \nabla_{\mathbf{X}} P_{c}=[X_{j}\partial P_{c}/\partial X_{j}]$, where $P_{c}$ is defined in Main Article \cref{softmax}. According to preliminary experiments, utilizing the gradient of the Softmax score ($\mathbf{X} \odot \nabla_{\mathbf{X}} P_{c}$) was more numerically stable than using the gradient of $\eta_{c}$ or even $\tilde{\eta}_{c}$ (the quantity explained by the Selective ISNet). Moreover, its results were similar to those obtained with the more time-consuming approach of creating one heatmap per logit ($\mathbf{X} \odot \nabla_{\mathbf{X}} y_{c}$, where $y_{c}$ is a classifier logit). The ISNet Softmax Grad*Input represents a new ablation experiment, where again the ISNet's LRP heatmaps were substituted by Gradient*Input explanations. However, here we consider an ablation model with a training computational efficiency similar to the Faster ISNet's.

\section{Data Processing and Augmentation}
\label{DataProcessing}

MNIST images were loaded in grayscale, standardized with pixel values between 0 and 1, and fed to the neural networks. In MNIST, we do not employ data augmentation, maximizing the synthetic bias influence over the classifier. Standard augmentation procedures (e.g., rotations and translations) can partially or totally remove the bias from the network input. The MNIST images' size is 28x28. The grayscale images were repeated in 3 channels to match the ResNet18 input shape. Although single-channel figures could be used, without profound changes in the classifier architecture they would provide minimal performance advantages over the standard 3-channel images. In Stanford Dogs, we simply loaded the RGB images, standardized their values between 0 and 1, and resized them to 224x224. Once more, we utilized no data augmentation, maximizing the strength of the synthetic bias, thus creating the most challenging scenario for the Faster ISNet.

The X-rays were processed and augmented following the procedures in the original ISNet study\cite{ISNet}, allowing direct comparisons to the neural networks trained in \cite{ISNet}. We loaded the images in grayscale, applied histogram equalization, resized them to 224x224 pixels, repeated the images in three channels, and performed data augmentation. Test images were made square by adding black rectangles to their borders before resizing. The strategy prevents any bias related to aspect ratio from affecting the test results\cite{ISNet}. By not employing the bars in training images, we do not allow neural networks to learn how to analyze the rectangles (especially the models without segmentation). Data augmentation was used only for training images, and it consisted in random translation (up to 28 pixels up/down and left/right), random rotation (from -40 to 40 degrees), and random flipping (50\% probability). Data augmentation was employed to avoid overfitting, and to make the DNN more resistant to the natural occurrence of the operations (rotations, translations and flipping). Moreover, it allows the Faster ISNet evaluation under a standard data augmentation protocol.

\section{Training Procedure}
\label{TrainingProcedure}

For a fairer comparison, we trained all neural networks with similar procedures and hyper-parameters. All models were implemented in PyTorch and trained on an NVIDIA RTX 3080 or an NVIDIA Tesla V100. All networks' trainable parameters were randomly initialized, because the utilization of pretrained models could reduce the generality of the conclusions drawn from a study assessing shortcut learning and resistance to background bias (refer to the original ISNet article for a detailed explanation\cite{ISNet}). Gradient norm clipping (with norm 1) was employed to improve the training procedure's stability. Weight decay was not utilized and the LRP-$\varepsilon$ $\varepsilon$ hyper-parameter was set to 0.01, like in the original ISNet study\cite{ISNet}. LRP-z$^{B}$ was employed for the first classifier layer in all ISNets, except for the Selective and Stochastic ISNets in the dog breed classification task. In this case, we empirically verified that the trained models' heatmaps were more interpretable and natural when LRP-z$^{B}$ was not used during training. The tuning procedure for loss hyper-parameters (in ISNets, multi-task U-Net, RRR and GAIN) followed the recipe in the original ISNet study\cite{ISNet}, except for the parameters $C_{1}$ and $C_{2}$, which were tuned with the novel technique in \cref{tuneCut}. The recipe in \cite{ISNet} is based on training DNNs with synthetically biased datasets and evaluating them on a validation dataset without the synthetic background bias. Synthetic bias allows one to simulate an o.o.d. evaluation dataset, even when the training and validation databases are actually i.i.d. upon removal of the synthetic bias. O.o.d. evaluation (real or simulated) is necessary for tuning hyper-parameters that are critical for hindering shortcut learning, such as ISNet's P and d.

For MNIST we utilized the stochastic gradient descend (SGD) optimizer, with momentum of 0.9, mini-batches of 128 samples, learning rate of 0.001, and 100 epochs. Hold-out validation was utilized to choose the best neural network after training. Most ISNet variations employed the loss hyper-parameters of: P=0.9, d=0.9, and $w_{1}=w_{2}=1$, but the Dual ISNet used P=0.95. For RRR, we set the Right Reasons loss' weight\cite{RRR} to 1000 (higher values led to a significant accuracy drop).

In dog breed classification, we employed SGD, mini-batches of 8 to 16 samples, learning rate of 0.001, and we trained for 200 epochs with hold-out validation. SGD momentum of 0.9 was utilized in GAIN and RRR. All other networks employed momentum of 0.99. Most Faster ISNets utilized the loss hyper-parameter d=0.999, except for the Dual ISNet, which used d=1. The Dual and Selective ISNets employed P=0.9, and the Stochastic ISNet P=0.85 without LDS and P=0.8 with LDS. The ISNet Softmax Grad*Input employed P=0.8 and d=0.999. Moreover, for all ISNets we set $w_{1}=1$ and $w_{2}=3$ (as in \cite{ISNet}), and did not tune these two parameters. RRR's Right Reasons loss' weight\cite{RRR} was 1000. We define the Multi-task U-Net loss as $L=P.L_{S}+(1-P).L_{C}$, where $L_{S}$ is the segmentation loss and $L_{C}$ the classification loss, both constituting cross-entropy. The parameter P balances the two terms, and we tuned it to P=0.99. GAIN loss weights were 1 for the classification loss, 1 for attention mining, and 100 for the external supervision\cite{GAIN}. The U-Net in the segmentation-classification pipeline was pre-trained for 200 epochs in the dog breed classification dataset, utilizing mini-batches of 16 samples, SGD with momentum of 0.99, hold-out validation, and learning rate of 0.001. It achieved 0.824 test intersection-over-union (IoU), using a segmentation mask threshold of 0.4 (defined to maximize validation IoU). In LRP Deep Supervision, individual heatmap losses are aggregated with GWRP\cite{GWRP} into a scalar heatmap loss, and we set GWRP's exponential decay parameter to 0.7. In the ResNet backbones, LDS supervises the heatmaps at the DNN input, at the global average pooling layer input, and at the input of each of the four ResNet stages\cite{ResNet}.

For COVID-19 detection, we closely followed the training procedure in the original ISNet paper\cite{ISNet}, as it guided the optimization of benchmark DNNs we compared the Faster ISNet to. Specifically, we trained for 48 epochs, with hold-out validation, learning rate of 0.001, stochastic gradient descend, momentum of 0.99, and mini-batches of 16 samples. All Faster ISNets employed the loss hyper-parameters d=0.999, P=0.7, $w_{1}=1$ and $w_{2}=3$. The ISNet Softmax Grad*Input used P=0.3 and d=0.999. Since the remaining benchmark networks are models trained in the original ISNet study\cite{ISNet}, their training hyper-parameters are described in \cite{ISNet}. 

\section{Statistical Analysis of Performance Metrics}
\label{statistical}

Performance metrics for MNIST and Stanford Dogs are normally reported as point estimates. Conversely, COVID-19 detection is a contemporary medical application. Thus, it is important to report uncertainty in the task. For COVID-19 detection, we present the DNNs' precision, recall, sensitivity, and F1-Score. The metrics are derived from the neural networks' confusion matrix. To provide their interval estimates we employ Bayesian evaluation. We resort to the same Bayesian model\cite{bayesianEstimator} that was used in the original ISNet study\cite{ISNet}. It was proposed to take a confusion matrix and output posterior estimates of precision, recall, and F1-Score. Moreover, it was expanded in \cite{bassi2021covid19} to also create interval estimates for specificity. Posteriors were estimated with Markov chain Monte Carlo (MCMC), implemented with the Python library PyMC3\cite{pymc}. We utilized the No-U-Turn Sampler\cite{NUTS}, using 4 chains and 110000 samples, being the first 10000 for tuning. The estimator's Monte Carlo error remained below 10$^{-4}$ for all metrics. We employed the same priors as the Bayesian model creators\cite{bayesianEstimator}. To calculate the macro-averaged ROC AUC (area under the receiver operating characteristic curve), we utilized a pair-wise approach\cite{MulticlassAUC} developed for multi-class single-label problems (such as COVID-19 classification). There is no established methodology to calculate interval estimated for this metric, and its authors suggested bootstrapping\cite{MulticlassAUC}, a technique that would not be feasible with the deep models and cross-dataset evaluation used in this study. 

\section{Datasets}
\label{datasets}

The datasets in this study simulate scenarios where background bias is present, creating a tendency for shortcut learning. There are two synthetically biased datasets: MNIST, a simple and popular database for computer vision, with low-resolution grayscale images; and Stanford Dogs, a dataset of high-resolution natural images, portraying a difficult fine-grained classification task\cite{StanfordDogs}. Furthermore, the COVID-19 X-ray dataset\cite{ISNet} represents a contemporary biomedical classification task, where non-synthetic background bias commonly undermines DNN generalization capacity\cite{ShortcutCovid,NatureCovidBias,ISNet,bassi2021covid19}.

\subsection{Synthetically Biased MNIST}

The biased MNIST dataset samples were drawn from the original MNIST database\cite{mnist}, a balanced dataset of hand-written digits (from 0 to 9), consisting in 60000 28x28 grayscale images. We consider the same test partition as the original database\cite{mnist} (10000 images), and randomly split the remaining 50000 images, selecting 80\% for training and 20\% for hold-out validation. 

\begin{figure}[h]
\includegraphics[width=0.2\textwidth]{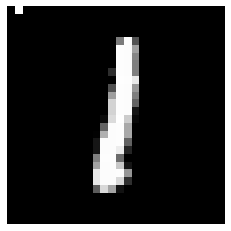}
\centering
\caption{Example of training sample from biased MNIST. The figure represents the digit 1. Accordingly, the second pixel (from left to right) in the image's top row was set to 1, representing background bias.}
\label{1MNIST}
\end{figure}

We added synthetic bias to all training data. The MNIST images' background is black (0). Consider an image displaying the digit c. To create a simple and attention-grabbing background bias, we set to white the (c+1)$^{th}$ pixel, from left to right, in the image's top row; e.g., for an image displaying the digit 0, we set the pixel in the top left corner to 1 (white). \Cref{1MNIST} provides an example of a training image. Due to its simplicity, it should be much easier for a classifier to produce decision rules based on the bias, instead of analyzing the more complex features of the hand-written digit. Therefore, we designed the dataset to be strongly prone to shortcut learning.

We also add the artificial bias to the hold-out validation dataset, which will be used to select the best performing DNN iteration after training. However, we employ an unbiased version of the validation set to tune the ISNet loss hyper-parameters P and d. Finally, we have three different test databases: the i.i.d. set has the same artificial bias as the training data; the o.o.d. dataset does not have any artificial bias; and the deceiving bias dataset has synthetic bias, but the correspondence between the ground-truth class and the highlighted background pixel is changed. In the deceiving bias dataset, we highlight the (c+2)$^{th}$ top pixel for the digit c, except for the digit 9, for which we highlight the first pixel (top left corner). A classifier suffering from shortcut learning should perform well on the i.i.d. test, lose accuracy on the o.o.d. test, and lose even more performance on the deceiving evaluation.

\subsection{Synthetically Biased Stanford Dogs}
The Stanford Dogs dataset\cite{StanfordDogs} is a subset of the ImageNet\cite{imagenet} database. It represents the task of dog breed classification, a fine-grained classification problem. The dataset portrays 120 different dog breeds, with 150 to 200 images per class, totaling 20580 samples. All images are RGB, with a resolution of at least 200x200. We utilize the testing data suggested by the dataset authors, which leave 100 images per class for training and at least 50 for evaluation\cite{StanfordDogs}. Moreover, we randomly split the training data, selecting 20\% of the samples for hold-out validation, and 80\% for training. All images are accompanied by ground-truth bounding boxes indicating the dogs. To increase task difficulty, we present the entire images to the classification neural networks, instead of cropping them according to the bounding boxes. The original ISNet study conducted experiments with a subset of the Stanford Dogs dataset (also with synthetic bias), consisting of only 3 classes and 300 training images \cite{ISNet}.

The dog breed classification task in Stanford Dogs was designed to be difficult\cite{StanfordDogs}. First, it presents a fine-grained classification task. Thus, dissimilarities between images of different classes can be small (little inter-class variation). Meanwhile, differences between samples of the same class may be large (large intra-class variation). Additionally, dogs display a wide variety of poses, ages, occlusion or self-occlusion, and color. Finally, backgrounds are complex and varied, including man-made environments. The baseline mean accuracy for the dataset, according to their authors\cite{StanfordDogs}, was only 22\%, even when considering just features from inside the images' bounding boxes.

To train the ISNet and many benchmark neural networks used in this study, we must have ground-truth segmentation masks indicating the images' region of interest, i.e., the dogs. We create them, from the available bounding boxes, utilizing DeepMAC, a pre-trained general purpose semantic segmenter, whose architecture was designed for novel class segmentation (segmenting objects whose class did not appear in the segmenter's training dataset)\cite{deepMAC}. The pre-trained DeepMAC is openly available for download\cite{deepMAC}. Upon visual evaluation, we confirmed that the automatically produced segmentation targets could precisely segment the dogs' bodies.

\begin{figure}[!h]
\includegraphics[width=0.2\textwidth]{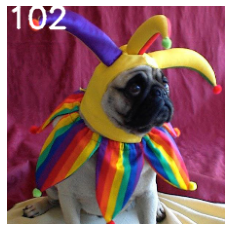}
\centering
\caption{Example of training sample from biased Stanford Dogs. The figure represents the Pug breed, identified by the number 102 in the training dataset (synthetic bias).}
\label{DogImage}
\end{figure}

We have added synthetic background bias to all Stanford Dogs training and hold-out validation samples. Again, we just keep an unbiased version of the hold-out validation dataset for tuning loss hyper-parameters. Like in MNIST, we design a background biasing feature whose identification is easier than the solution of the desired classification task (i.e., identification the dog breed). Having 120 classes, we associated each one to a number, from 0 to 119, and inserted this number in the images' top left corner. The number for a class is simply the class label, which is the same label proposed in the original Stanford Dogs dataset, minus one (the original labels ranged from 1 to 120, not 0 to 119). \Cref{DogImage} displays an image from the training dataset. Naturally, identifying a typed number is simpler than dog breed classification, especially considering that the numbers have a regular placement, font, color (white), and size (10\% of the image height). Like in the biased MNIST, we create three diverse test datasets. The i.i.d. test set has the same bias as the training data, the o.o.d. set has no synthetic bias, and the deceiving bias test has a synthetic bias designed to fool biased classifiers. The deceiving bias is a random number, from 0 to 119, which has no digit in common with the number that represented the test image's class during training; e.g., the class Chihuahua is associated with the number 0 during training. Thus, in the deceiving bias dataset, Chihuahua pictures will have a random number from 1 to 119, except for any number containing a digit 0.

\subsection{COVID-19 X-ray Classification}

The COVID-19 X-ray dataset represents high-resolution data (224x224) and exemplifies a natural occurrence of background bias. It is common for large X-ray databases portraying a disease not to have enough samples representing other conditions of interest, or to have such samples drawn from diverse hospitals. To train reliable deep classifiers, large databases adequately representing all conditions of interest are important. When such data is not available from a single source (or from a set of sources with similar class distributions), studies normally employ mixed source databases, where samples representing different categories (e.g., diseases) were drawn from distinct locations (e.g., hospitals). Each source can be correlated to particular background characteristics (e.g., specific text in the X-ray background\cite{FirstPaperCovid}). Since, in mixed datasets, the different sources represent different classes, their distinct background features become correlated to the image labels, constituting background bias and causing shortcut learning\cite{NatureCovidBias,ShortcutCovid,ISNet}.

Dataset mixing is common in COVID-19 classification\cite{reviewCovid}. The COVID-19 dataset from the original ISNet study\cite{ISNet} purposefully exemplifies such scenario. Each class (COVID-19, pneumonia and healthy) was extracted from a different dataset. These datasets originated from diverse sets of hospitals and cities. They were chosen due to their large size, high quality and adequate documentation. Hold-out validation images were randomly selected, and they share sources with the training images. Hold-out validation represents 25\% of the non-test images. There is no patient with X-rays in the training and validation splits. Importantly, to estimate generalization to real-world data, the dataset test samples were not randomly drawn from the same sources providing the training and hold-out validation data. Instead, we utilize an out-of-distribution test dataset. I.e., all test X-rays were drawn from hospitals and cities never utilized to gather training or hold-out validation data. For a detailed description of the dataset, including information about data acquisition time and methodology, disease severity, exclusion criteria, equipment used, data format, and labeling scheme, please refer to the study presenting the database\cite{ISNet}. In the remainder of this Section, we briefly summarize it.

All images in the dataset are frontal X-rays from adult patients, collected from open sources. Specifically, the COVID-19 training and hold-out validation data consists in 4644 positive samples from the Brixia COVID-19 X-ray dataset\cite{BrixiaSet}, being all images that show visible (Brixia score higher than 0) signs of the disease. All images come from the same hospital, ASST Spedali Civili di Brescia, Brescia, Italy. The database is one of the largest COVID-19 data collections to date, and it was created to simulate a real clinical scenario: the images represent all triage and patient monitoring (in sub-intensive and intensive care units) X-rays collected between March 4th and April 4th, 2020, reflecting real-world variability\cite{BrixiaSet}. Training and validation X-ray images showing the healthy (4644) and pneumonia (4644) conditions were randomly extracted from the CheXPert database, one of the largest and most well-known chest X-ray datasets\cite{irvin2019chexpert}. This data collection originated in the Stanford University Hospital, California, United States of America\cite{irvin2019chexpert}. COVID-19 images were labeled by radiologists on-shift\cite{BrixiaSet}. Meanwhile, the other classes were labeled by natural language processing, according to radiological reports\cite{irvin2019chexpert}.

The external (o.o.d.) test dataset has 1515 COVID-19 X-rays, extracted from the BIMCV COVID-19+ (iterations 1+2)\cite{BimcvSet} database. The data collection is also among the largest open COVID-19 X-ray datasets. It was produced between February 16th and June 8th, 2020. Samples were obtained from the health departments in the Valencian healthcare system, Spain, which involves multiple hospitals, mostly in the provinces of Alicante and Castellón\cite{BimcvSet}. Furthermore, our o.o.d. test data contains 1295 test X-rays representing the pneumonia class, which were gathered from the ChestX-ray14 database\cite{chex14}. Like CheXPert, the data collection is also one of the largest and most well known chest X-ray databases. It was created in the National Institutes of Health Clinical Center, Bethesda, United States of America. Finally, the healthy class X-rays were extracted from a database assembled in Montgomery County, Maryland, USA (80 images), and Shenzhen, China (326 images)\cite{ChineseDataset1}. They were gathered from the Montgomery County’s Tuberculosis screening program, in collaboration with the local Department of Health and Human Services, and from Shenzhen No.3 People’s Hospital, Guangdong Medical College (Shenzhen, China). All the healthy class samples were manually labeled by radiologists\cite{ChineseDataset1}. COVID-19 images were labeled by radiologists and correspond to patients with at least one positive COVID-19 PCR or immunological test in the BIMCV COVID-19+ data acquisition period (February 16th to June 8th, 2020)\cite{BimcvSet}. Meanwhile, the pneumonia labels were assigned by natural language processing, according to radiological reports\cite{chex14}.

\subsubsection{X-ray Dataset Limitations.} 
The X-ray dataset we employ was built in \cite{ISNet} to represent mixed COVID-19 X-ray databases, which were the norm for AI training during most of the COVID-19 pandemic \cite{NatureCovidBias,reviewCovid}. As background bias and shortcut learning were common in such datasets \cite{NatureCovidBias}, the motivation behind this X-ray database is promoting research on the consequences of dataset mixing in a contemporary medical task\cite{ISNet}. Accordingly, the database is not the ideal choice for one trying to create a diagnostic AI model for real-world use, due to limitations listed in \cite{ISNet} and summarized below. The creation of a diagnostic model for real-word use is not the objective of our study. We use COVID-19 detection as a demonstrative task, to assess the Faster ISNet robustness to background bias. Without clinical tests, we do not claim diagnostic performance, nor point our models as substitutes for RT-PCR exams. 

The dataset's first limitation is that it has only adult patients. Because the COVID-19 data had much less pediatric patients than the other classes, such patients were removed to avoid bias. Second, all COVID-19 samples are from 2020. Thus, the dataset does not represent newer variants of the disease, nor vaccinated patients. Classification performance may be different for current cases. Finally, the database may under-represent asymptomatic cases of COVID-19 and people that did not seek medical help, as all COVID-19 and pneumonia images came from a hospital environment. Additionally, radiologists found signals of COVID-19 in all dataset's COVID-19 X-rays\cite{ISNet}. This characteristic improves training and avoids false positives, but the database under-represents mild cases or cases with radiological manifestations that are very hard to detect \cite{ISNet}.

\section{Training and Run-time Speed}
\label{speed}

The Faster ISNet and the original ISNet have the same run-time speed as a standard classifier because the creation of LRP heatmaps is not necessary after training; e.g., any ISNet with a DenseNet121 backbone will be as fast as a standard DenseNet121 at run-time. In COVID-19 detection, the segmentation-classification pipeline (U-Net followed by DenseNet121) was the second-best performing neural network, after the ISNet. Moreover, in the original ISNet study\cite{ISNet}, it was the only benchmark model that could reliably avoid the influence of synthetic background bias on the classifier's decisions, like the ISNet. However, at run-time, the model is much slower than any ISNet variant, as it involves running two deep neural networks, instead of one. The original ISNet study compared the run-time ISNet with a DenseNet121\cite{DenseNet} backbone to the pipeline composed of a U-Net\cite{unet} and the DenseNet121. Considering mixed-precision, mini-batches of 10 images (of size 224x224x3), and an NVIDIA RTX 3080, the ISNet was able to process an average of 353 samples per second, while the pipeline processed 207. Without mixed-precision, the ISNet processed 298 samples per second, and the pipeline 143. Thus, the ISNet was 70 to 108\% faster\cite{ISNet}. Moreover, the model had only 8M parameters, while the pipeline had 39M, being about 5 times larger. Therefore, the original and the Faster ISNets are significantly faster and lighter than the pipeline after training. Such efficiency is especially important for its deployment in mobile or less capable devices.

The original ISNet's training time increases approximately linearly with the number of classes in the classification problem\cite{ISNet}. Considering a classification problem with 3 classes, 24183 training images and 2993 validation samples, the original ISNet took about 1300 s per epoch\cite{ISNet}. Meanwhile, the benchmark neural networks training times were: 1500 s for GAIN, 900 s for RRR, 320 s for the segmentation-classification pipeline (and 400 s per epoch when pre-training its U-Net), 240 s for the DenseNet121, 250 s for the multi-task U-Net, 160 seconds for the vision transformer, and 60 s for the AG-Sononet\cite{ISNet}. 

\begin{figure}[!h]
\includegraphics[width=0.7\textwidth]{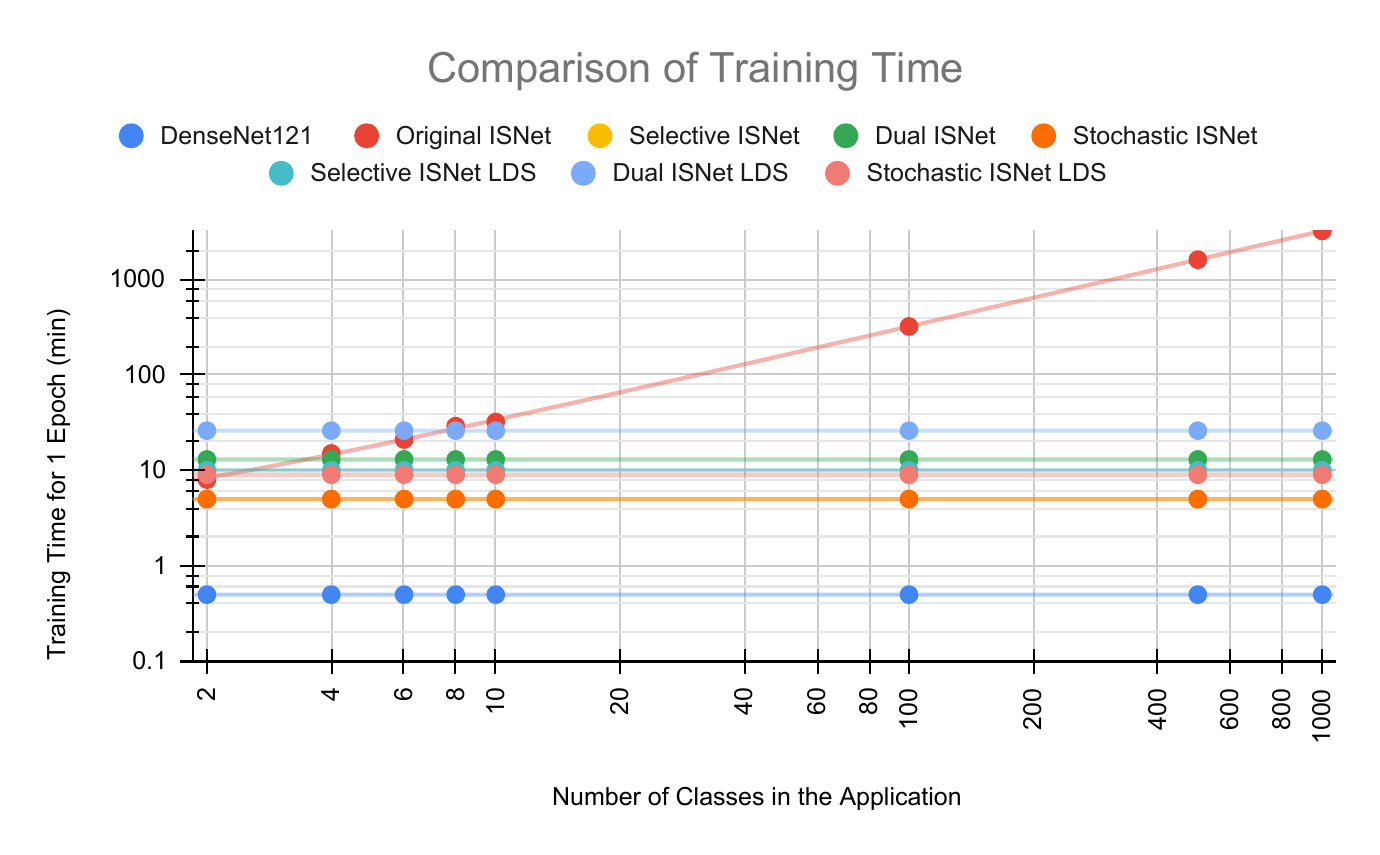}
\centering
\caption{Training time for multiple neural networks, considering one epoch of 13932 224x224x3 images, and a varying number of categories in the classification problem.}
\label{TrainingTime}
\end{figure}

The Faster ISNet's training time does not increase linearly with the number of classes in the classification task. \Cref{TrainingTime} compares the training time for the original and the Faster ISNets, employing a log-log plot. It displays the time for one epoch, considering 13932 images (with 80\% used for training and 20\% for validation), with resolution of 224x224 and 3 channels. The networks ran in an NVIDIA RTX 3080 with mixed precision. We simulated 1 to 1000 classes in the classification task, by varying the number of output neurons in the networks. For realism, we set a fixed limit of GPU (graphics processing unit) memory; 10 gigabytes, which is about the limit in the RTX 3080. Accordingly, the mini-batch size is always set to approximately utilize the 10 gigabytes of video memory. The original ISNet memory utilization increases while we generate LRP heatmaps in parallel, thus we reduce batch size while increasing the number of classes. When batch size reaches one, we start generating heatmaps in series, which makes the model's training time increase linearly. In \cref{TrainingTime}, the original ISNet's batch size reaches one with about 10 classes (considering 10 GB of video memory).

As expected, the original ISNet training time increases approximately linearly with the number of classes in the classification task. Meanwhile, the training time for all Faster ISNet variations (Selective, Dual, and Stochastic) is independent of the number of classes, like for the standard DenseNet121. The common classifier training time was 0.5 min per epoch, the Stochastic ISNet and the Selective ISNet achieved both 5 min per epoch, and the Dual ISNet, needing to produce two LRP heatmaps per image, required 13 minutes per epoch. With LRP Deep Supervision (LDS), the Stochastic ISNet, Selective ISNet, and Dual ISNet needed 9, 10 and 26 minutes per epoch, respectively. Considering the networks without LDS, the original ISNet was always slower than the Selective and Stochastic ISNets. It was faster than the Dual ISNet until about 3 classes in the classification problem, becoming increasingly slower afterwards. To put the numbers in \cref{TrainingTime} into perspective, it would take about 16 hours to train the Selective or Stochastic ISNet for 200 epochs (considering the data and hardware we used to generate \cref{TrainingTime}). Meanwhile, the original ISNet would require 26 hours for 2 classes, 100 hours for 10 classes, and about 440 days with 1000 classes. Its training time could be reduced with more powerful hardware and parallelism over multiple GPUs. However, such settings represent higher cost and electrical energy consumption. Therefore, besides improving training time, the Faster ISNet expands the number of applications where LRP optimization can be feasibly implemented.

\section{Heatmap Analysis}
\label{heatmaps}

\begin{figure}[!h]
\includegraphics[width=0.4\textwidth]{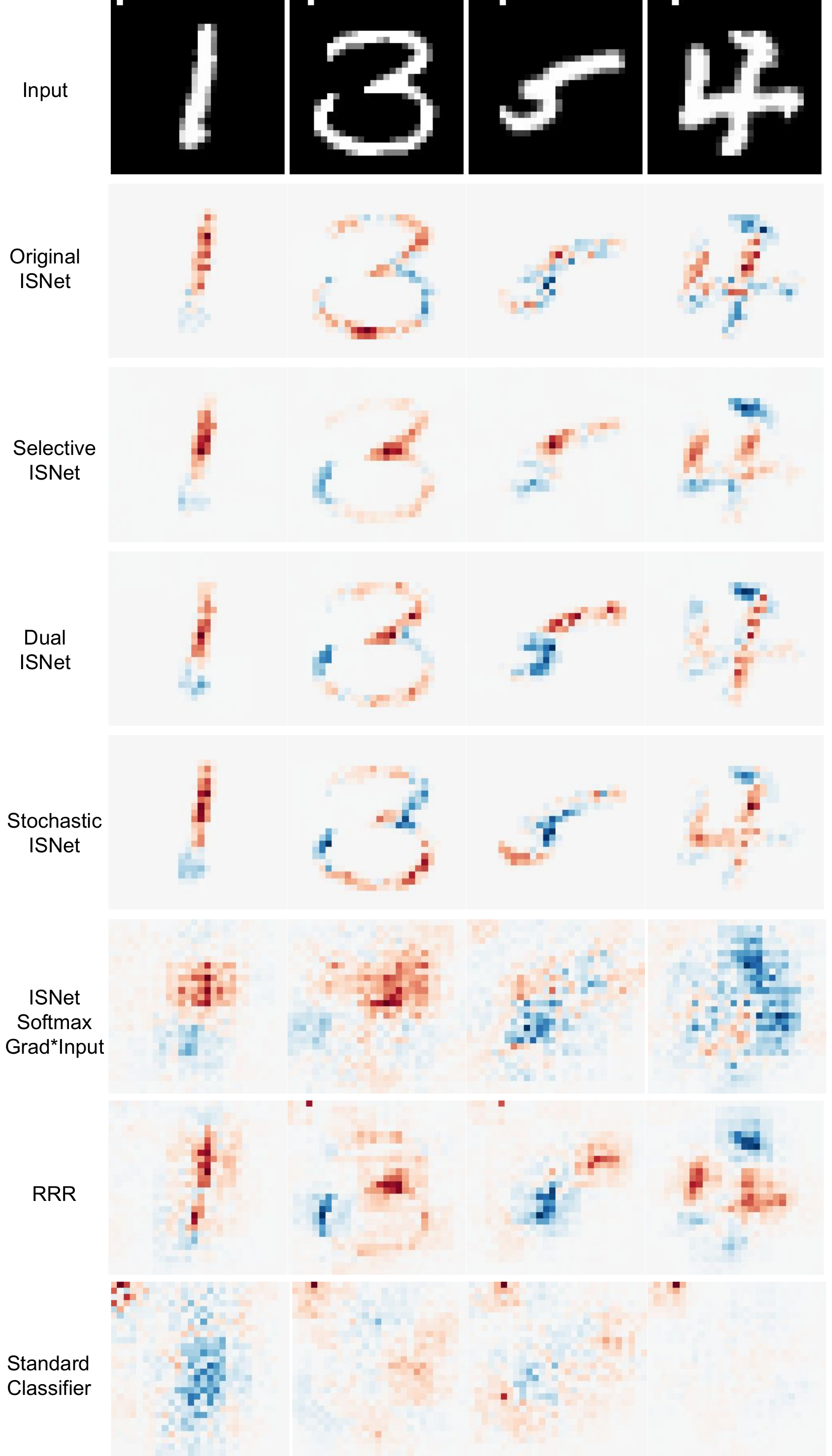}
\centering
\caption{LRP heatmaps for the MNIST classification task. A white pixel in the images' top left corner represents synthetic background bias. Red colors indicate areas that contributed for the DNN confidence in the images' true class, while blue areas reduced such confidence. White regions were not important for the DNN decision.}
\label{MNISTMaps}
\end{figure}

\Cref{MNISTMaps} presents LRP heatmaps for biased MNIST test images (i.i.d. test). Thus, the images contain the same synthetic background bias found in training data (the white pixels in the images' upper region). In this Section, all LRP heatmaps were created with LRP-$\varepsilon$ throughout the DNN, and LRP-$z^{B}$ for the input layer. Red colors (positive relevance) indicate areas that the DNN associated with the ground-truth class. Negative relevance is shown in blue, representing regions that decreased the classifier's confidence in the ground-truth class. White regions were not important for the classifier's decisions (low absolute relevance). \Cref{MNISTMaps} heatmaps demonstrate that the bias could not influence the Faster and Original ISNets. The ISNet Softmax Grad*Input shows minimal bias attention, only in the image displaying a 5. RRR paid some attention to bias in 2 of the 4 images, while the standard classifier mostly focused on bias. Such findings corroborate with the quantitative results in Main Article \cref{AllResults}, which prove that the synthetic bias had no influence over the Faster and Original ISNet decisions, it had a small influence over RRR, and major influence over the standard ResNet18. Although Main Article \cref{AllResults} shows no accuracy drop upon bias removal for the ISNet Softmax Grad*Input, there is a very small drop when we analyze results with one additional significant figure: its accuracy was 0.949 in i.i.d. testing, 0.9486 in o.o.d., and 0.9485 in confounding bias. The 0.04-0.05\% accuracy drop explains the tiny attention to bias the network demonstrates in \cref{MNISTMaps}.

\begin{figure}[!h]
\includegraphics[width=0.525\textwidth]{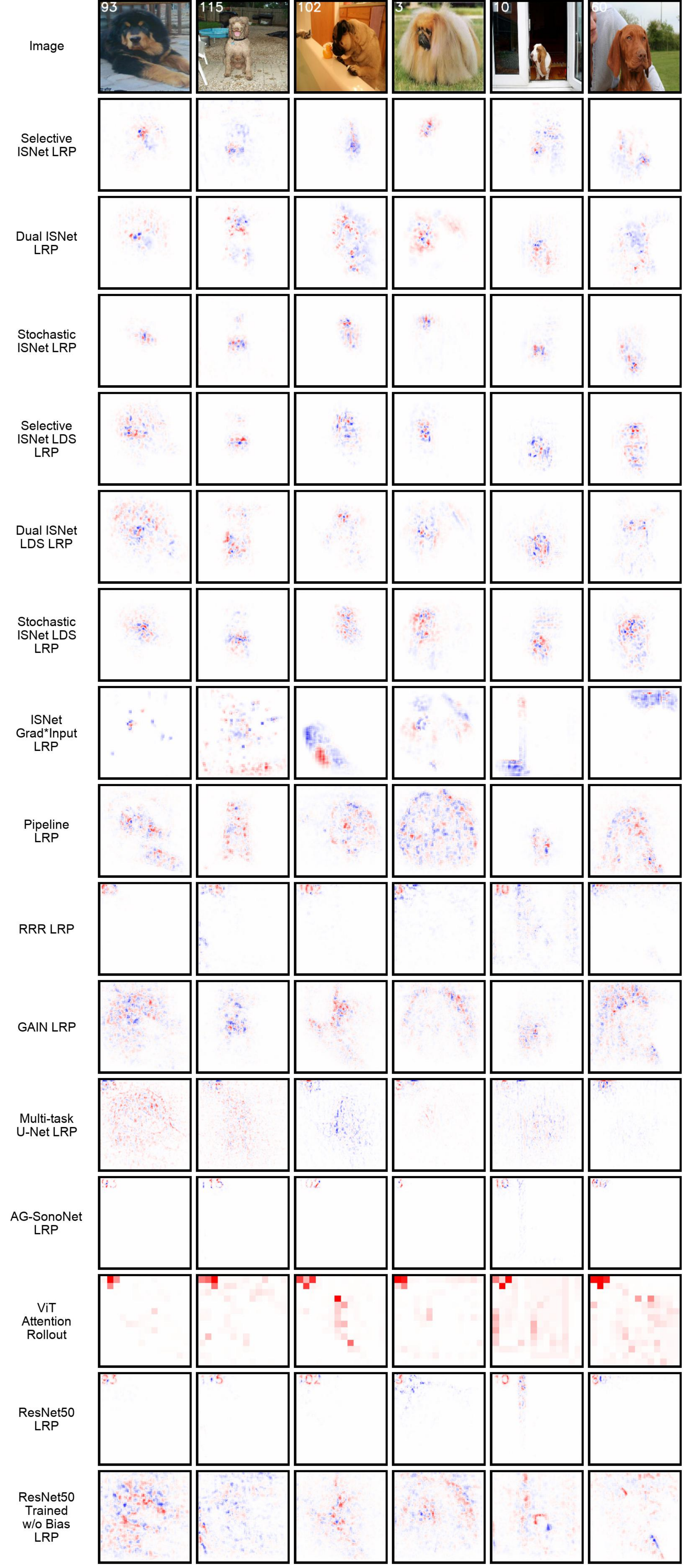}
\centering
\caption{Heatmaps for the Stanford Dogs classification task. Numbers in the images' top left corner represent synthetic background bias. Red colors indicate areas that contributed for the DNN confidence in the images' true class, while blue areas reduced such confidence. White regions were not important for the DNN decision.}
\label{DOGSMaps}
\end{figure}

\Cref{DOGSMaps} presents the heatmaps for the synthetically biased Stanford Dogs i.i.d. test dataset. Numbers in the images' top left corner indicate the dogs' breeds, representing synthetic background bias. All neural networks are explained with LRP, except for the Vision Transformer, which is explained with attention rollout, the standard explanation methodology for the architecture\cite{VisionTransformer}. Attention rollout does not differentiate positive and negative attention, and it is not class selective. However, it can clearly show that the Vision Transformer paid attention to the background bias. \Cref{DOGSMaps} demonstrates that, except for the Faster ISNet variants (S. \cref{DOGSMaps} rows 2 to 7), the segmentation-classification pipeline (row 9), and GAIN (row 11), all other neural networks paid significant attention to the numbers in the figures' backgrounds. Therefore, the heatmaps in the figure corroborate with the numerical results in Main Article \cref{AllResults}, which quantitatively prove that the three aforementioned models were the only ones whose decisions were not influenced by background bias. For reference, the last row in the figure presents attention maps for the ResNet50 trained and tested without synthetic background bias.

GAIN could minimize shortcut learning, and it did not achieve a spurious mapping solution in Stanford Dogs. Spurious mapping\cite{ISNet} happens when a classifier learns to produce deceiving GradCAM heatmaps, by mapping features form the background of an image to the foreground of the late DNN feature maps. This undesirable solution allows a DNN to minimize GAIN's losses, while still having decision rules that consider background bias. In the paper introducing the ISNet, this problem appeared when GAIN was trained for multiple tasks that considered background bias, including the COVID-19 detection problem that we also consider in this work\cite{ISNet}. Accordingly, spurious mapping is a possible but not guaranteed outcome of Grad-CAM optimization.

Main Article \cref{heatmapsFig1} portrays LRP heatmaps for the COVID-19 classification task. Please refer to the original ISNet study \cite{ISNet} for heatmaps representing the remaining neural networks in Main Article \cref{AllResults} (RRR, GAIN, multi-task DNN, AG-Sononet, vision transformer, and segmentation-classification pipeline). In summary, the heatmaps in Main Article \cref{heatmapsFig1} and in \cite{ISNet} demonstrate that all neural networks have clear attention to background bias in COVID-19 detection, except for the segmentation-classification pipeline and the LRP-based ISNets (original, Selective, Dual, and Stochastic ISNet). Here, background means all X-ray regions outside of the lungs. A reduction of background bias' influence over the classifier indicates a reduction of shortcut learning, resulting in superior o.o.d. test performance\cite{ShortcutLearning}. Accordingly, the LRP-based ISNets significantly surpassed the remaining neural networks on the external (o.o.d.) COVID-19 test dataset (Main Article \cref{AllResults}). Hence, the quantitative results in Main Article \cref{AllResults} support the information in the heatmaps presented in Main Article \cref{heatmapsFig1} and in the original ISNet study\cite{ISNet}. Overall, LRP heatmaps and numerical results (i.e., o.o.d. performance and accuracy drop upon synthetic bias removal) show that the LRP-based ISNets and the segmentation-classification pipeline are the only implemented models that were consistently resistant to background bias in all experiments in this study and in \cite{ISNet}. However, the pipeline may pay attention to the erased background's shape and to the foreground's borders, forming decision rules that represent another form of shortcut learning \cite{ISNet}. 
 

\section{Supplementary Results}
\label{moreResults}
\subsection{Supplementary Tables}
\Cref{additionalResultsTable} presents AUC\cite{DeLongAUC} and F1-Scores (per class and average) for the external COVID-19 detection database. F1-Scores are presented as mean$\pm$std, [HDI]. Mean refers to the metric's mean value, considering its probability distribution, and std refers to its standard deviation. The 95\% high density interval (HDI) is an interval containing 95\% of the metric's probability mass. Moreover, any point inside the interval has a higher probability density than any point outside of the interval. Please refer to Section \ref{statistical} for more details about our statistical analysis.

\begin{table}[!h]
\centering
\caption{Additional o.o.d. evaluation scores for COVID-19 detection}
\label{additionalResultsTable}
\begin{tabular}{|l|l|l|l|l|l|} 
\hline
\begin{tabular}[c]{@{}l@{}}Neural \\Network\end{tabular}          & \begin{tabular}[c]{@{}l@{}}Normal \\F1-Score\end{tabular}                & \begin{tabular}[c]{@{}l@{}}Pneumonia \\F1-Score\end{tabular}             & \begin{tabular}[c]{@{}l@{}}COVID-19 \\F1-Score\end{tabular}              & \begin{tabular}[c]{@{}l@{}}Macro-average \\F1-Score\end{tabular}         & \begin{tabular}[c]{@{}l@{}}Macro \\OvO \\AUC\end{tabular}  \\ 
\hline
\begin{tabular}[c]{@{}l@{}}Selective\\ISNet\end{tabular}          & \begin{tabular}[c]{@{}l@{}}0.625$\pm$0.017,\\{[}0.59,0.658]\end{tabular}  & \begin{tabular}[c]{@{}l@{}}0.783$\pm$0.01,\\{[}0.764,0.802]\end{tabular}  & \begin{tabular}[c]{@{}l@{}}0.902$\pm$0.006,\\{[}0.891,0.913]\end{tabular} & \begin{tabular}[c]{@{}l@{}}0.77$\pm$0.008,\\{[}0.754,0.786]\end{tabular}  & 0.946                                                      \\ 
\hline
\begin{tabular}[c]{@{}l@{}}Dual \\ISNet\end{tabular}              & \begin{tabular}[c]{@{}l@{}}0.534$\pm$0.018,\\{[}0.498,0.57]\end{tabular}  & \begin{tabular}[c]{@{}l@{}}0.731$\pm$0.01,\\{[}0.71,0.751]\end{tabular}   & \begin{tabular}[c]{@{}l@{}}0.881$\pm$0.006,\\{[}0.868,0.893]\end{tabular} & \begin{tabular}[c]{@{}l@{}}0.715$\pm$0.009,\\{[}0.699,0.733]\end{tabular} & 0.909                                                      \\ 
\hline
\begin{tabular}[c]{@{}l@{}}Stochastic \\ISNet\end{tabular}        & \begin{tabular}[c]{@{}l@{}}0.604$\pm$0.018,\\{[}0.568,0.639]\end{tabular} & \begin{tabular}[c]{@{}l@{}}0.709$\pm$0.011,\\{[}0.687,0.731]\end{tabular} & \begin{tabular}[c]{@{}l@{}}0.879$\pm$0.006,\\{[}0.867,0.891]\end{tabular} & \begin{tabular}[c]{@{}l@{}}0.731$\pm$0.009,\\{[}0.714,0.748]\end{tabular} & 0.935                                                      \\ 
\hline
\begin{tabular}[c]{@{}l@{}}ISNet Softmax\\Grad*Input\end{tabular} & \begin{tabular}[c]{@{}l@{}}0.273$\pm$0.015,\\{[}0.244,0.302]\end{tabular} & \begin{tabular}[c]{@{}l@{}}0.09$\pm$0.01,\\{[}0.07,0.111]\end{tabular}    & \begin{tabular}[c]{@{}l@{}}0.604$\pm$0.01,\\{[}0.585,0.623]\end{tabular}  & \begin{tabular}[c]{@{}l@{}}0.323$\pm$0.007,\\{[}0.308,0.337]\end{tabular} & 0.608                                                      \\ 
\hline
\begin{tabular}[c]{@{}l@{}}Original \\ISNet\end{tabular}          & \begin{tabular}[c]{@{}l@{}}0.555$\pm$0.022,\\{[}0.512,0.597]\end{tabular} & \begin{tabular}[c]{@{}l@{}}0.858$\pm$0.007,\\{[}0.844,0.871]\end{tabular} & \begin{tabular}[c]{@{}l@{}}0.907$\pm$0.006,\\{[}0.896,0.918]\end{tabular} & \begin{tabular}[c]{@{}l@{}}0.773$\pm$0.009,\\{[}0.755,0.791]\end{tabular} & 0.952                                                      \\ 
\hline
\begin{tabular}[c]{@{}l@{}}Standard\\Classifier\end{tabular}      & \begin{tabular}[c]{@{}l@{}}0.444$\pm$0.02,\\{[}0.403,0.482]\end{tabular}  & \begin{tabular}[c]{@{}l@{}}0.434$\pm$0.015,\\{[}0.405,0.463]\end{tabular} & \begin{tabular}[c]{@{}l@{}}0.76$\pm$0.008,\\{[}0.744,0.775]\end{tabular}  & \begin{tabular}[c]{@{}l@{}}0.546$\pm$0.01,\\{[}0.527,0.565]\end{tabular}  & 0.808                                                      \\ 
\hline
\begin{tabular}[c]{@{}l@{}}U-Net+\\Classifier\end{tabular}        & \begin{tabular}[c]{@{}l@{}}0.571$\pm$0.018,\\{[}0.535,0.607]\end{tabular} & \begin{tabular}[c]{@{}l@{}}0.586$\pm$0.013,\\{[}0.561,0.611]\end{tabular} & \begin{tabular}[c]{@{}l@{}}0.776$\pm$0.008,\\{[}0.76,0.792]\end{tabular}  & \begin{tabular}[c]{@{}l@{}}0.645$\pm$0.009,\\{[}0.626,0.663]\end{tabular} & 0.833                                                      \\ 
\hline
\begin{tabular}[c]{@{}l@{}}Multi-task \\U-Net\end{tabular}        & \begin{tabular}[c]{@{}l@{}}0.419$\pm$0.025,\\{[}0.369,0.469]\end{tabular} & \begin{tabular}[c]{@{}l@{}}0.119$\pm$0.011,\\{[}0.098,0.14]\end{tabular}  & \begin{tabular}[c]{@{}l@{}}0.585$\pm$0.009,\\{[}0.566,0.602]\end{tabular} & \begin{tabular}[c]{@{}l@{}}0.374$\pm$0.01,\\{[}0.355,0.394]\end{tabular}  & 0.553                                                      \\ 
\hline
AG-Sononet                                                        & \begin{tabular}[c]{@{}l@{}}0.124$\pm$0.015,\\{[}0.096,0.153]\end{tabular} & \begin{tabular}[c]{@{}l@{}}0.284$\pm$0.015,\\{[}0.255,0.312]\end{tabular} & \begin{tabular}[c]{@{}l@{}}0.659$\pm$0.009,\\{[}0.641,0.676]\end{tabular} & \begin{tabular}[c]{@{}l@{}}0.356$\pm$0.008,\\{[}0.34,0.372]\end{tabular}  & 0.591                                                      \\ 
\hline
\begin{tabular}[c]{@{}l@{}}Extended\\GAIN\end{tabular}            & \begin{tabular}[c]{@{}l@{}}0.203$\pm$0.019,\\{[}0.166,0.24]\end{tabular}  & \begin{tabular}[c]{@{}l@{}}0.485$\pm$0.013,\\{[}0.46,0.511]\end{tabular}  & \begin{tabular}[c]{@{}l@{}}0.711$\pm$0.009,\\{[}0.693,0.728]\end{tabular} & \begin{tabular}[c]{@{}l@{}}0.466$\pm$0.009,\\{[}0.449,0.485]\end{tabular} & 0.724                                                      \\ 
\hline
RRR                                                               & \begin{tabular}[c]{@{}l@{}}0.36$\pm$0.018,\\{[}0.325,0.394]\end{tabular}  & \begin{tabular}[c]{@{}l@{}}0.552$\pm$0.013,\\{[}0.526,0.577]\end{tabular} & \begin{tabular}[c]{@{}l@{}}0.737$\pm$0.009,\\{[}0.72,0.755]\end{tabular}  & \begin{tabular}[c]{@{}l@{}}0.55$\pm$0.009,\\{[}0.532,0.568]\end{tabular}  & 0.775                                                      \\ 
\hline
\begin{tabular}[c]{@{}l@{}}Vision \\Transformer\end{tabular}      & \begin{tabular}[c]{@{}l@{}}0.382$\pm$0.017,\\{[}0.348,0.415]\end{tabular} & \begin{tabular}[c]{@{}l@{}}0.474$\pm$0.013,\\{[}0.448,0.499]\end{tabular} & \begin{tabular}[c]{@{}l@{}}0.525$\pm$0.011,\\{[}0.503,0.548]\end{tabular} & \begin{tabular}[c]{@{}l@{}}0.46$\pm$0.009,\\{[}0.443,0.478]\end{tabular}  & 0.683                                                      \\
\hline
\end{tabular}
\end{table}

\subsection{Supplementary Analysis of Results}
The original ISNet study justified why the ISNet surpassed the benchmark DNNs\cite{ISNet}. In summary, the classifier in the pipeline could pay attention to the erased background and to the foreground's borders, indicating biased decision rules, which consider the foreground shape and position. Such rules can impair o.o.d. generalization. The multi-task DNN could accurately segment the foreground, but its classification scores were influenced by background bias. Therefore, the DNN's attention foci for the segmentation and classification tasks diverged. Standard attention mechanisms, which do not learn from segmentation targets (e.g., Vision Transformer and AG-Sononet), cannot reliably differentiate important foreground features from background bias. Thus, they may consider the bias relevant and not reduce shortcut learning\cite{ISNet}. Input gradients and Gradient*Input create noisy explanations for deep neural networks. Accordingly, with respect to the optimization of these alternative explanation techniques (RRR\cite{RRR} and ISNet Grad*Input), the optimization of LRP-$\varepsilon$ (ISNet) better and more stably converged when using very deep backbones (DenseNet121 or ResNet50) and large images (224x224), leading to superior resistance to background bias and o.o.d. accuracy\cite{ISNet}. Finally, DNNs that optimized Grad-CAM (e.g., GAIN\cite{GAIN}) could learn to produce deceiving Grad-CAM heatmaps, which showed no background attention for classifiers whose decisions were being influenced by background bias (spurious mapping)\cite{ISNet}. The phenomenon is inherent to the Grad-CAM formulation, and it cannot affect LRP optimization\cite{ISNet}. 

The results in this study confirm the analysis in \cite{ISNet}. The Faster ISNets significantly surpassed the segmentation-classification pipeline in COVID-19 detection, a task where the model displayed attention to foreground borders and erased backgrounds\cite{ISNet}. This background attention was less noticeable in Stanford Dogs (\cref{DOGSMaps}), where the pipeline could match the ISNets. A similar result was observed and analyzed for a subset (only 3 classes) of Stanford Dogs in \cite{ISNet}. 

In Stanford Dogs, the multi-task U-Net could also precisely segment the foreground (dogs), achieving 0.824 test intersection-over-union (IoU). However, its outputs were influenced by synthetic bias (Main Article \cref{AllResults}). The same phenomenon was observed for all experiments in \cite{ISNet}.

The Vision Transformer and AG-Sononet considered the synthetic bias in Stanford Dogs, as the networks did in \cite{ISNet}. Consistently, the AG-Sononet was the network with the strongest focus on synthetic background bias in Stanford Dogs and in all experiments from \cite{ISNet}. The network has the largest accuracy drop upon bias removal in Main Article \cref{AllResults}, and its LRP heatmaps display the strongest concentration of relevance on bias (\cref{DOGSMaps}).

The optimization of input gradients (RRR) and Gradient*Input could only lead to high accuracy and robustness to background bias when DNNs were shallower, i.e., using a ResNet18 backbone in our experiments and a VGG-19 in \cite{ISNet}. The methods were inaccurate and/or biased in all experiments using deeper backbones (ResNet50 or DenseNet121), both in this study and in \cite{ISNet}.

Finally, \cite{ISNet} demonstrates that spurious mapping is a possible outcome of Grad-CAM optimization, representing a shortcut solution to Grad-CAM-based losses. Dataset, model, and training dynamics can define whether the DNN arrives at this solution or not\cite{ISNet}. Accordingly, GAIN did not suffer spurious mapping in Stanford Dogs, but it arrived at the shortcut solution multiple times in \cite{ISNet} (in COVID-19 detection, tuberculosis detection, and even dog breed classification with a subset of Stanford Dogs).

\end{document}